\documentclass[notitlepage,12pt,tightenlines,nofootinbib,showkeys]{revtex4-1}
\pdfoutput=1

\usepackage{amsmath}
\usepackage{amssymb,amsfonts}
\usepackage{bm}
\usepackage[mathcal]{euscript}
\usepackage{graphicx}
\usepackage{subfig}
\usepackage{verbatim}
\usepackage{ifthen}
\usepackage{natbib}

\graphicspath{{figs/}{figs_lo/}}

\newcommand{\COLOR}{true}
\newcommand{\ifcolor}[2]{\ifthenelse{\equal{\COLOR}{true}}{#1}{#2}}
\ifcolor{\usepackage{color}}{}

\newcommand{\coloronline}[2]{\ifcolor{#1}{#2 [#1 online]}}

\newcommand{\DRAFT}{false}
\newcommand{\ifdraft}[2]{\ifthenelse{\equal{\DRAFT}{true}}{#1}{#2}}

\ifdraft{%
  \ifcolor{%
  \newcommand{\jlt}[1]{{\color{red} [JLT: #1]}}
  \newcommand{\mra}[1]{{\color{blue} [MRA: #1]}}
  \newcommand{\ch}[1]{{\color{cyan} #1}} % mark changes
  \newcommand{\delete}[1]{} % deleted text
  }{%
  \newcommand{\jlt}[1]{{\sl [JLT: #1]}}
  \newcommand{\mra}[1]{{\sl [MRA: #1]}}
  \newcommand{\ch}[1]{{\sl #1}} % mark changes
  \newcommand{\delete}[1]{} % deleted text
  }
}{%
  \newcommand{\jlt}[1]{}
  \newcommand{\mra}[1]{}
  \newcommand{\ch}[1]{}
  \newcommand{\delete}[1]{} % deleted text
}

\newlength{\figwidth} % reference width for figures
\newcommand{\mathnotation}[2]{\newcommand{#1}{\ensuremath{#2}}}

\newcommand{\eg}{\textit{e.g.}}

\mathnotation{\ldef}{\mathrel{\raisebox{.069ex}{:}\!\!=}}% Left define
\mathnotation{\rdef}{\mathrel{=\!\!\raisebox{.069ex}{:}}}% Right define
\mathnotation{\tim}{t}				% Time
\mathnotation{\nn}{n}				% # of strands
\mathnotation{\Br}{B}				% Braid group
\mathnotation{\R}{R}				% Domain
\mathnotation{\ip}{i}				% Counter for braid elements
\mathnotation{\jp}{j}				% Counter for braid elements
\mathnotation{\per}{m}				% iterates of periodic orbit
\mathnotation{\htop}{h}				% Topological entropy
\mathnotation{\htopf}{\htop_{\mathrm{flow}}}	% Topological entropy of flow
\mathnotation{\ac}{a}				% Dynnikov coord a
\mathnotation{\bc}{b}				% Dynnikov coord b
\mathnotation{\cc}{c}				% Dynnikov coord c
\mathnotation{\dc}{d}				% Dynnikov coord d
\mathnotation{\fc}{f}				% Dynnikov coord f
\mathnotation{\acnew}{\ac'}
\mathnotation{\bcnew}{\bc'}
\mathnotation{\abv}{\bm{u}}			% Dynnikov coord vector
\mathnotation{\Nint}{L}				% Intersection #
\mathnotation{\Nreal}{N_{\mathrm{real}}}	% Number of realizations
\mathnotation{\dt}{\Delta\tim}			% Length of time intervals
\mathnotation{\ti}{q}				% Label of time intervals
\mathnotation{\tmax}{\tim_{\mathrm{max}}}	% Max integration time

\newcommand{\LCS}{Lagrangian coherent structure}
\newcommand{\LCSs}{{\LCS}s}
\newcommand{\IPS}{invariant puncture set}
\newcommand{\IPSs}{{\IPS}s}
\newcommand{\EPS}{entangled puncture set}
\newcommand{\EPSs}{{\EPS}s}
\newcommand{\ES}{entangled set} % short form of EPS
\newcommand{\ESs}{{\ES}s}

\begin{document}

\title{Detecting coherent structures using braids}

\author{Michael R. Allshouse}
\email{allshouse@mit.edu}
\affiliation{Department of Mechanical Engineering, MIT, Cambridge, MA 02139,
  USA}
\author{Jean-Luc Thiffeault}
\email{jeanluc@math.wisc.edu}
\affiliation{Department of Mathematics, University
  of Wisconsin, Madison, WI 53706, USA}

\date{11 June 2011}


\begin{abstract}
  The detection of coherent structures is an important problem in
  fluid dynamics, particularly in geophysical applications.  For
  instance, knowledge of how regions of fluid are isolated from each
  other allows prediction of the ultimate fate of oil spills.
  Existing methods detect Lagrangian coherent structures, which are
  barriers to transport, by examining the stretching field as given by
  finite-time Lyapunov exponents.  These methods are very effective when
  the velocity field is well-determined, but in many applications only
  a small number of flow trajectories are known, for example when
  dealing with oceanic float data.  We introduce a topological method
  for detecting invariant regions based on a small set of
  trajectories.  In the method we regard the two-dimensional
  trajectory data as a braid in three dimensions, with time being the
  third coordinate.  Invariant regions then correspond to trajectories
  that travel together and do not entangle 
  other trajectories.  We detect these regions by examining the growth 
  of hypothetical loops surrounding sets of trajectories, and searching 
  for loops that show negligible growth.
\end{abstract}

\maketitle

\section{Introduction}
\label{sec:intro}

The ability to accurately identify regions of mixing and barriers to
transport in two-dimensional systems has applications in the
ocean~\cite{BeronVera2008, Mezic2010}, the
atmosphere~\cite{Nakamura1996, Shuckburgh2003}, mantle
flows~\cite{Farnetani2003}, as well as granular flows
\cite{Drake1990}.  Central to this is the concept of \LCSs, which are
barriers to transport between different dynamical regions in a flow.
Thus, fluid in a region delimited by \LCSs\ does not mix well with the
surrounding fluid.  These structures can be impossible to detect in
the Eulerian (fixed) frame, since they move around and change their
shape, possibly in an irregular manner.  Over the last decade, new
methods were developed to find \LCSs~\cite{Haller2000b, Haller2001,
  Haller2001c, Haller2002, Mathur2007, Mezic2010, Tang2010,
  Haller2011}.  All of these methods require field data (\eg\ velocity
or vorticity fields) which in many practical situations is not readily
accessible.  Instead, the data available is often in the form of
\emph{tracer trajectories}, either from oceanic floats or atmospheric
balloons.  Though it may contain the trajectory of dozens of floats,
such data is too sparse to reconstruct a velocity field.  A method
which estimates the location of \LCSs\ based on trajectory data is
thus highly desirable.

A first step towards this was taken by~\citet{Thiffeault2005,
  Thiffeault2010}, who regarded a set of two-dimensional float
trajectories as a \emph{braid} in three dimensions, where the third
dimension is time.  This is a well-established technique for studying
two-dimensional trajectories~\cite{Gambaudo1999, Boyland2000,
  Boyland2003, Vikhansky2003, Thiffeault2005, Thiffeault2006, Kin2005,
  Gouillart2006, Nechaev, Thiffeault2010, Turner2011}.
\citet{Thiffeault2010} applied recently-developed mathematical
techniques~\cite{Moussafir2006} to the rapid computation of the
\emph{topological entropy} of trajectory datasets.  The
topological entropy is related to the degree of \emph{entanglement} of
trajectories: a set of trajectories trapped in a vortex, for instance,
would have zero topological entropy.  A measure of chaos, and hence of
mixing, was thus obtained without requiring the full velocity field.
\citet{MattFinn2007} showed numerically that as the number of
trajectories is increased, the topological entropy associated with the
trajectories approaches the true topological entropy of the underlying
flow~\cite{Fathi1979, Thurston1988, Newhouse1993, Boyland1994}.

In the present paper, we implement the next step --- the detection of
invariant regions based on sparse particle trajectory data.  These
regions contain fluid that possibly mixes well with other fluid within
the region, but not with fluid outside.  The regions can move and
change shape aperiodically, and each region has its own topological
entropy.

If a set of trajectories are within such an invariant region, then
from the braid perspective they form a `coherent bundle.'  A coherent
bundle can be thought of as rope, which is made up of small strands
(the individual trajectories).  We will detect such bundles in the
following manner.  Consider an imaginary material loop drawn around
a bundle of trajectories.  We say that the loop is `pulled-tight' if
it is tightened to its shortest length that still encloses the
trajectories (see Fig.~\ref{fig:loop}).  If the bundle really is
inside an invariant region, then the length of the pulled-tight loop
does not grow rapidly in time.  On the other hand, if any trajectory
is outside of a region, then typically the chaotic action of the flow
will cause this loop to grow exponentially, even if it is
pulled-tight.

What allows this calculation to be practical is that from the
topological perspective we do not actually need to follow true
material loops in the fluid.  Rather, there is a symbolic description
of pulled-tight loops that allows a very rapid computation of their
evolution, in particular the growth rate of their length.  These are
relatively recent topological tools which are only now being applied
to physical problems~\cite{Moussafir2006,Thiffeault2010}.

We must emphasize that the detection method presented here is not
perfect: like all finite-time approaches, it is susceptible to the
trajectory data being too short, or too sparse --- for instance, it is
not possible to reliably detect an invariant region based on a single
trajectory.  But when relatively few trajectories are known, our
approach probably yields most --- if not all --- of the topological
information that is theoretically available, though this has not been
shown rigorously.

\begin{figure*}
  \centering
  \subfloat[]{\label{fig:sys1}\includegraphics[width=0.4\figwidth]{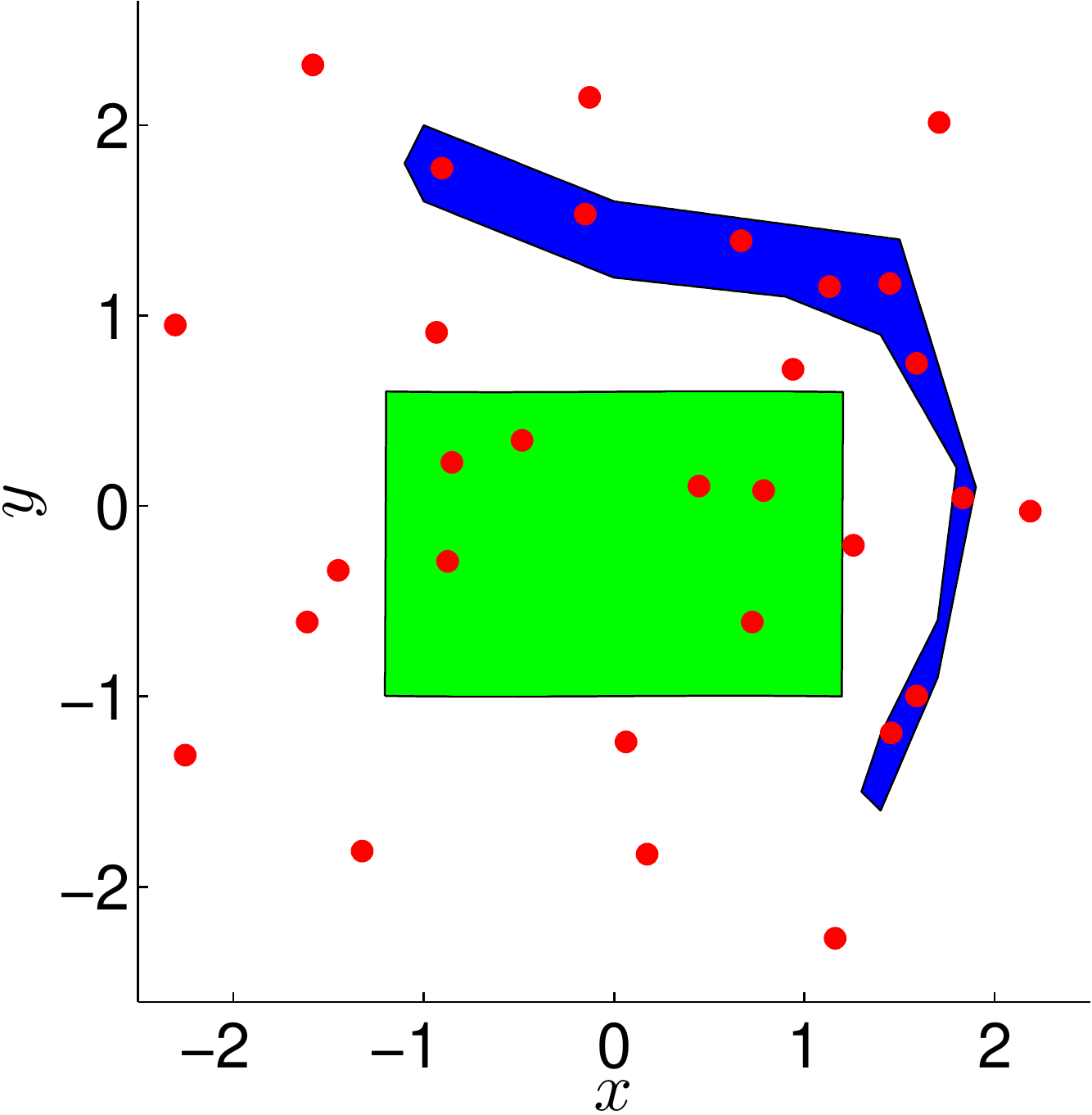}}
  \hspace{.03\figwidth}
  \subfloat[]{\label{fig:sys2}\includegraphics[width=0.4\figwidth]{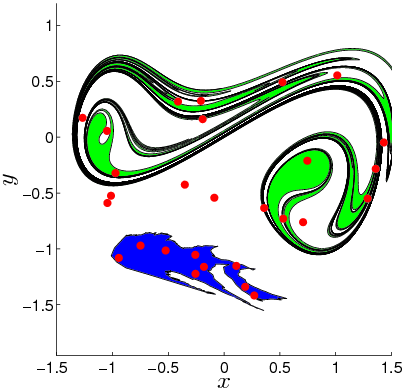}}     
  \caption{(a) Initial state and (b) state after evolution by the
    Duffing system of Section~\ref{sec:sys}.  The
    \protect\coloronline{blue}{dark} region is surrounded by a
    transport barrier, while the \protect\coloronline{green}{light}
    region is not.  Particle positions are shown as
    \protect\coloronline{red dots}{dots}.}
  \label{fig:sys}
\end{figure*}

Throughout this paper, we use a modified version of the Duffing
oscillator to test and illustrate the features of the method.  A
typical time-evolution of this model system is depicted in
Fig.~\ref{fig:sys}: Fig.~\ref{fig:sys}a shows 30 trajectories as well
as two regions of interest (\coloronline{blue and green}{light and
  dark}) in an initial state.  These are shown at a later time in
Fig.~\ref{fig:sys2}.
While the boundaries of the two regions are distorted, the
trajectories that start within a region remain in that region for all
time.  The \coloronline{blue}{dark} region has changed shape,
but its pulled-tight perimeter has remained approximately the same
length.  This indicates that the boundary of the
\coloronline{blue}{dark} region would be classified as a potential
transport barrier (\LCS).  The \coloronline{green}{light} region shows
substantial folding and stretching of its boundary as trajectories
from outside crisscross the region.  The boundary of the
\coloronline{green}{light} region thus tends to grow exponentially,
even if it is pulled-tight around the trajectories.  The
boundary of the \coloronline{green}{light} region would then not be
considered a transport barrier.  In this paper, we will show how to
automate the detection of loops which do not grow rapidly, such as the
one enclosing the \coloronline{blue}{dark} region.

The paper is divided as follows.  In Section~\ref{sec:braid}, we
introduce the necessary tools from braid theory and topological
surface dynamics.  We then demonstrate how to use these tools to
detect invariant regions in Section~\ref{sec:poc}, using a direct
approach of searching over a large set of loops.  In
Section~\ref{sec:alg}, we refine the method to make it much quicker,
by focusing on loops enclosing pairs of trajectories.  Both the direct
and the improved method are tested on the modified Duffing oscillator.
We test the refined approach on a two-dimensional rod-stirring device
in Section~\ref{sec:rod}, and offer some conclusions in
Section~\ref{sec:con}.  Because this paper contains some terminology
that may not be familiar to many readers, we have included a Glossary
in an appendix.

\section{Braid theory and topological surface dynamics}
\label{sec:braid}

In this section, we present a number of definitions and tools from
braid theory and topological dynamics.  These will be kept to a
minimum, and we give references containing a more complete treatment.
Specifically, we focus on applying the theory to a set of trajectories
evolving in time on a two-dimensional plane.  As a visual example, a
sample set of three particle trajectories is shown in
Fig.~\ref{fig:def1}.  The trajectories are constrained to two
dimensions in physical space. The particles are often referred to as
{\it punctures}, since they are regarded as topological obstacles
analogous to small holes in the plane.  For example, there are three
punctures in Fig.~\ref{fig:def1} which are represented by dots.  The
initial positions of the punctures are shown as crosses.

\subsection{Physical braids}
\label{sec:physbraids}

\begin{figure}
  \centering
  \subfloat[]{\label{fig:def1}\includegraphics[width=0.31\figwidth]{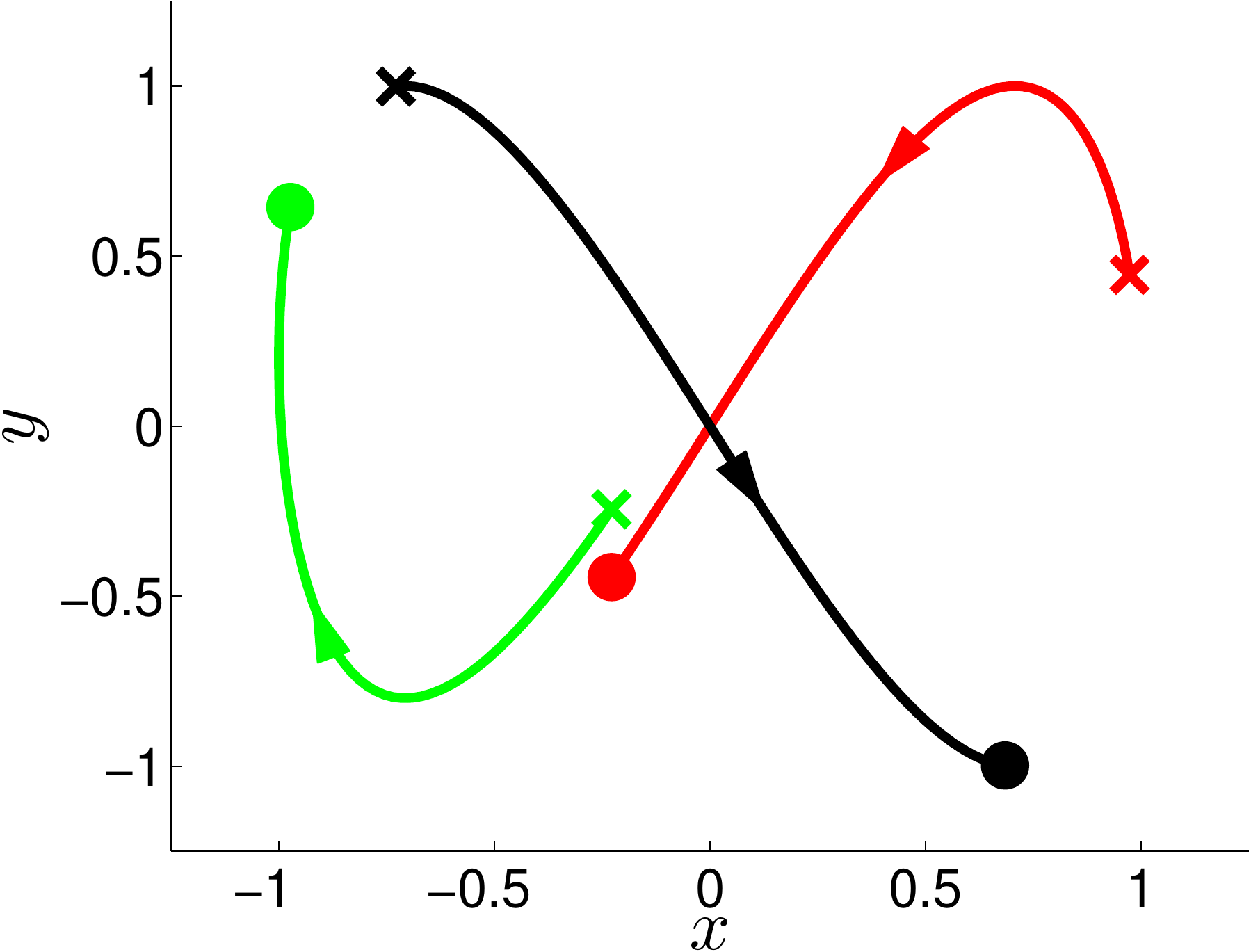}}
  \hspace{.02\figwidth}
  \subfloat[]{\label{fig:def2}\includegraphics[width=0.3\figwidth]{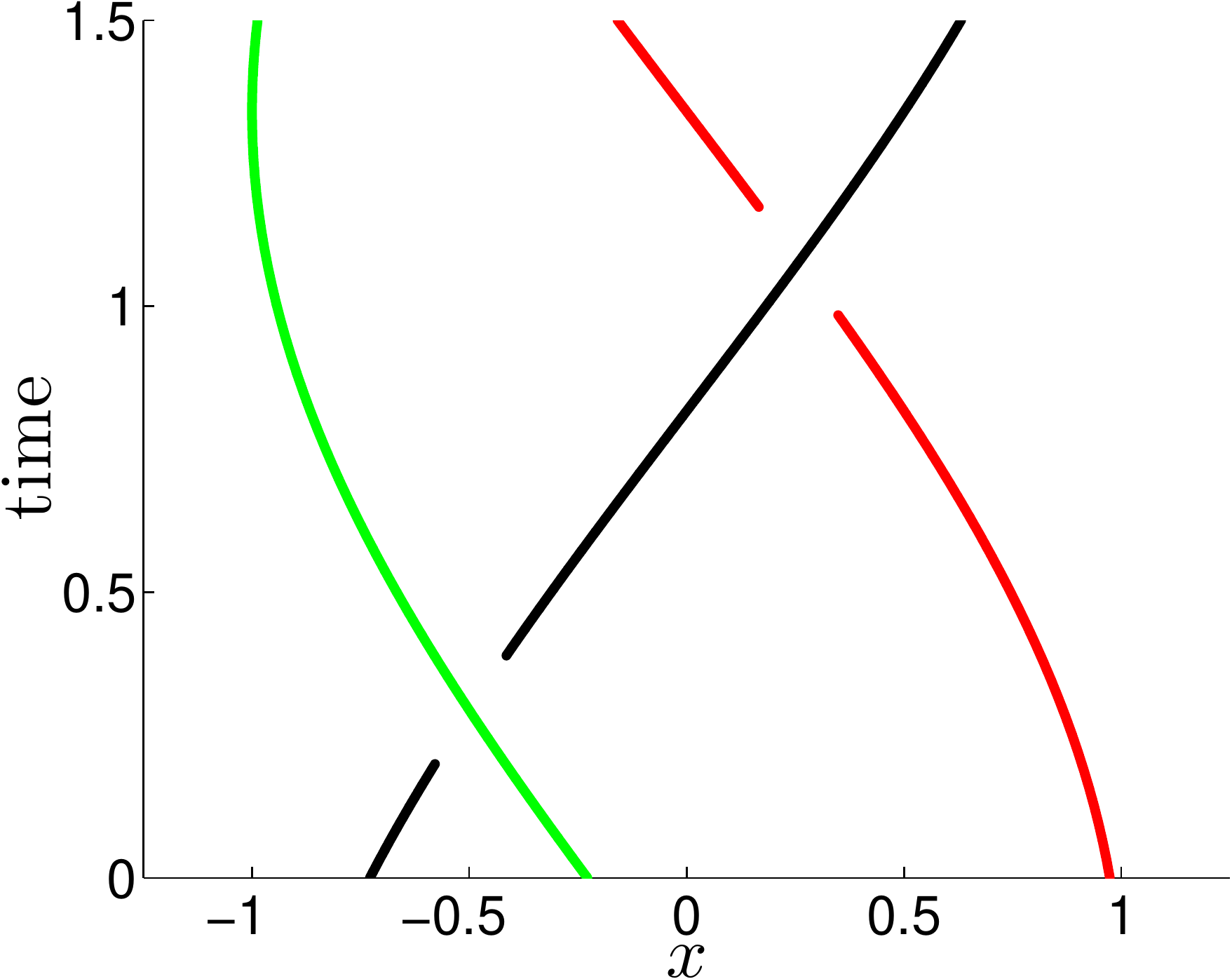}}
  \hspace{.02\figwidth}
  \subfloat[]{\label{fig:def4}\includegraphics[width=0.275\figwidth]{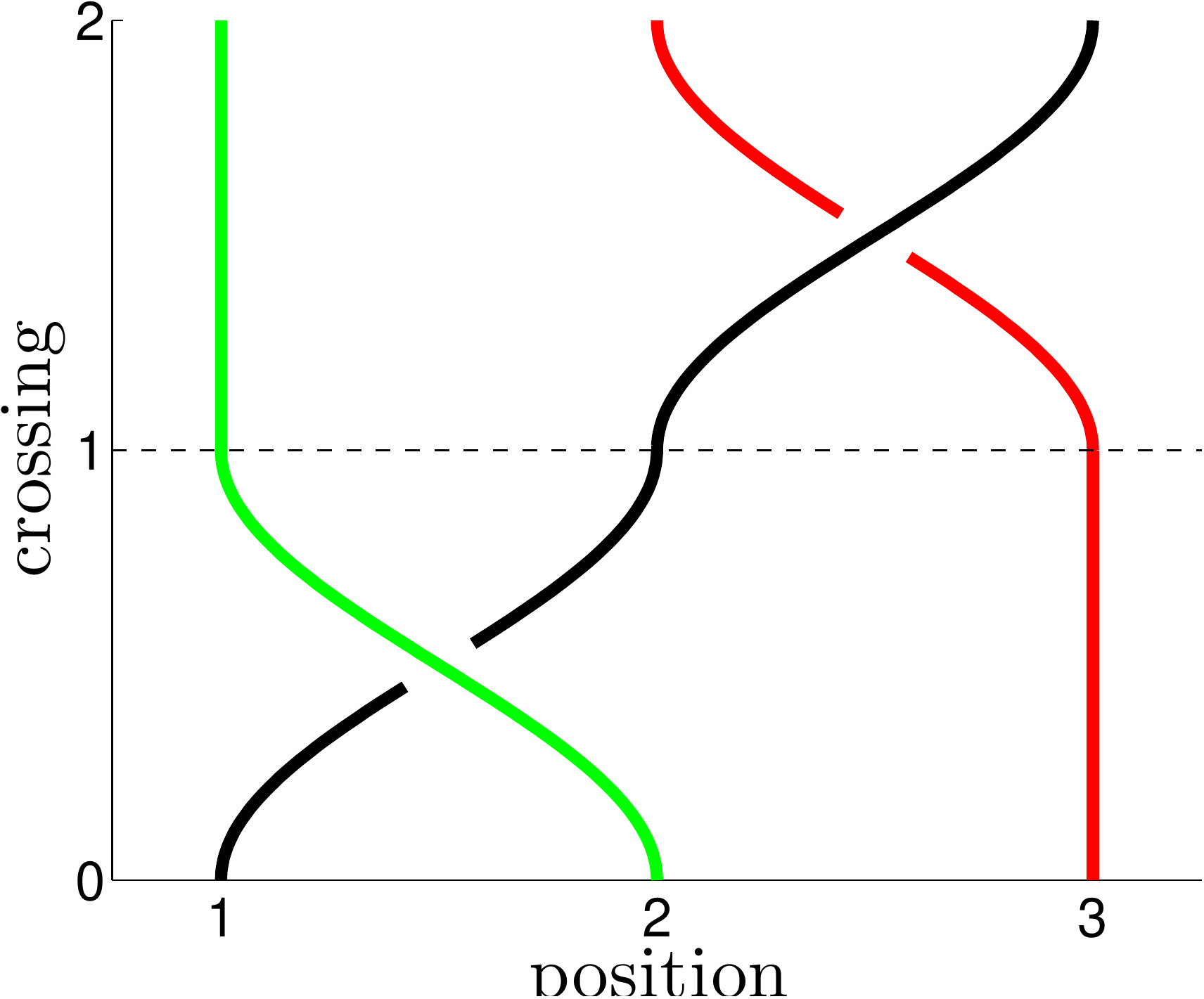}}
  \caption{(a) Three sample trajectories in the physical plane.  A
    cross represents the initial position of the trajectory and a dot
    the final position.  (b) The projection of the trajectories onto
    the $x$-axis as a function of time.  (c) The braid in (b) shown as
    a standard braid diagram.}
  \label{fig:def}
\end{figure}

The particle trajectories can be lifted from a two-dimensional domain
to a three-dimensional space where the vertical axis is time.  These
three-dimensional space-time trajectories are called {\it strands}.
Due to the monotonic nature of time, a strand can only move upwards,
since it cannot travel back in time.  If these three-dimensional
strands are projected onto the plane containing the $x$-axis and time
then Fig.~\ref{fig:def1} becomes
Fig.~\ref{fig:def2}.  Note that the crossing information was preserved
in making the projection, which records which strand was above another
in the~$y$ coordinate.  The collection of strands defines a {\it
  physical braid}.  Two physical braids are considered equivalent if
they can be continuously deformed into one another without the strands
crossing each other or boundaries.  Figure~\ref{fig:def4} is the
\emph{standard braid diagram} corresponding to Fig.~\ref{fig:def2}, in
which the strands were displaced into standardized `elementary
crossings,' with only one crossing occurring at a time.  On the
horizontal axis we record the left-to-right position of each particle
or puncture.

\subsection{Generators}
\label{sec:gen}

Figure~\ref{fig:def} shows that the topological information contained
in a set of trajectories --- who passes in front of who, and when ---
can be reduced to a fairly simple picture (Fig.~\ref{fig:def4}).  From
there it is straightforward to pass from this geometric description to
an \emph{algebraic} one, where crossings are written as a sequence of
symbols.  For instance, the braid in Fig.~\ref{fig:def4} is written
algebraically as
\begin{equation}
  \sigma_1\,\sigma_2^{-1}\,.
  \label{eq:def4braid}
\end{equation}
The symbol~$\sigma_1$ denotes the clockwise interchange of the first
and second particles (when viewed from above), and the
symbol~$\sigma_2^{-1}$ the counterclockwise interchange of the second
and third particles.  Note that first, second, etc.\ refer here to the
position of a particle from left to right along the~$x$-axis at any
point in time.  The individual symbols in~\eqref{eq:def4braid} are
ordered from left to right in time.  In general, $\sigma_i$ denotes
the clockwise interchange of the~$i$th and~$(i+1)$th particles,
with~$i$ ranging from~$1$ to~$\nn-1$ for~$\nn$ particles.  The
symbols~$\sigma_i$ are called \emph{generators of the braid group
  on~$\nn$ strands}, though we also refer to the~$\sigma_i^{-1}$ as
generators.  This is indeed a group, more precisely a
finitely-generated group with an infinite number of elements.  The
group-theoretic description of braids is due to~\citet{Artin1947}; see
the book by~\citet{Birman1975} or chapters in many books on knot
theory~\cite{Adams, Murasugi}.

Thus, the topological information in a set of~$\nn$ trajectories can
be written as a sequence of~$\nn-1$ generators and their inverses.
These generators are obtained by following the trajectories,
projecting along the~$x$-axis, and recording when two particles
exchange positions along the~$x$-axis.  We call this exchange a
crossing.  To determine the sign of the generator (i.e., whether it
is~$\sigma_i$ or~$\sigma_i^{-1}$), we look at the~$y$ coordinate of the
two particles at the time of crossing.  If the~$y$ coordinate of
the~$i$th particle is greater than that of the~$(i+1)$th particle the
generator is~$\sigma_i$; otherwise it is~$\sigma_i^{-1}$.  For a more
complete description of the process of extracting generators from
trajectories, see~\citet{Thiffeault2005,Thiffeault2010}.

\subsection{Loops}
\label{sec:loops}

\begin{figure}
  \centering
  \subfloat[]{\label{fig:physloop}
    \includegraphics[width=0.4\figwidth]{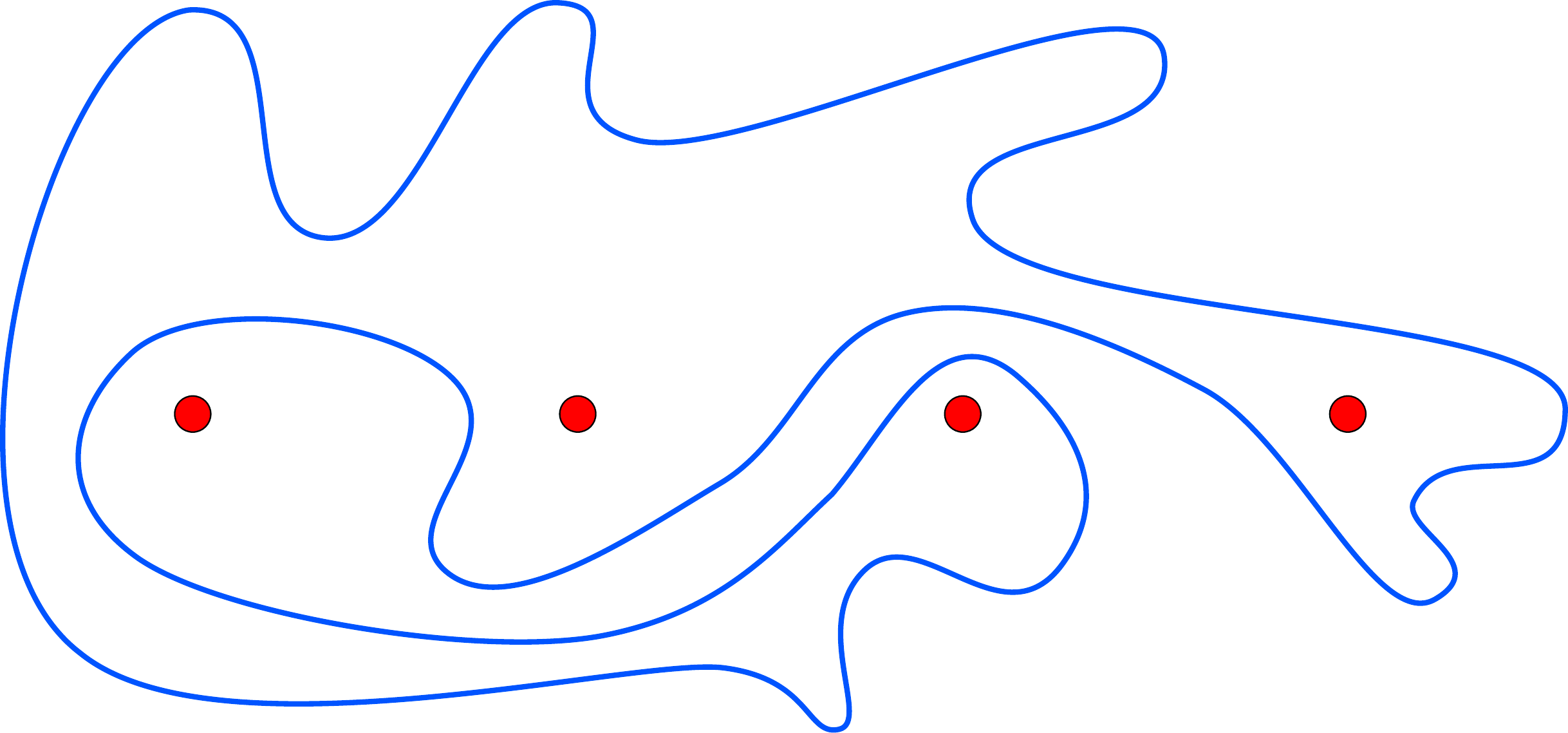}}
  \hspace{.04\figwidth}
  \subfloat[]{\label{fig:def5}\includegraphics[width=0.5\figwidth]{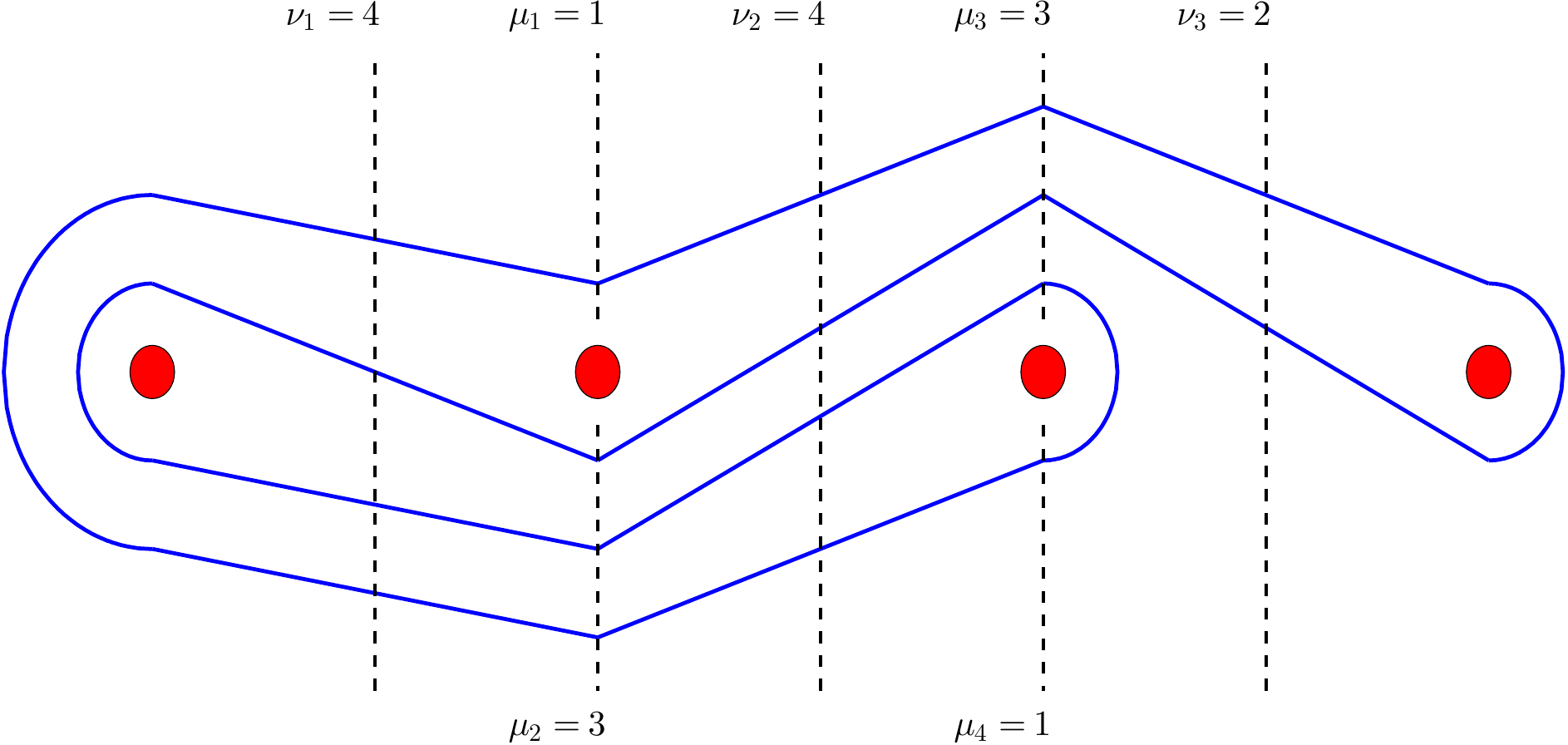}}
  \caption{(a) Physical loop around four punctures.  (b) Simplified
    (`pulled-tight') representation of the same loop, called a
    topological loop.  The crossing numbers $\mu_i$ and $\nu_i$ are
    also shown; the corresponding Dynnikov coordinates are $a = [1\
    -1]$ and $b = [0\ 1]$.}
  \label{fig:loop}
\end{figure}

So far we have shown how a set of particle trajectories can be
converted to a physical braid, and that physical braid can then in
turn be converted to an algebraic sequence of generators while
preserving the essential topological information.  Now we will pursue
an analogous path for~\emph{material loops}.  A sample loop is drawn
in Fig.~\ref{fig:physloop}, encapsulating four punctures (or
trajectories) depicted as disks.  We will only deal with loops that
are \emph{non-self-intersecting}, i.e., which do not cross themselves.
This is a natural property of material loops in fluids, since by
determinism two initially-distinct fluid particles can never
simultaneously occupy the same point in space.  We are concerned here
with \emph{topological loops} with respect to the particle
trajectories.  That is, two loops are considered topologically
identical if they can be deformed into each other without crossing any
punctures.  The process of passing from a physical material loop to a
simplified topological loop is analogous to passing from
Fig.~\ref{fig:def2} to Fig.~\ref{fig:def4} for braids.
Figure~\ref{fig:def5} shows the simplified topological loop
corresponding to Fig.~\ref{fig:physloop}, obtained by pulling-tight on
the loop without crossing the punctures.

Now that we have passed from physical material loops to simplified
topological loops, we can go a step further and provide a symbolic
description of loops.  The crucial fact is that any
non-self-intersecting closed topological loop, winding around
punctures, can be reconstructed merely by counting the number of
intersections with fixed reference lines, such as the seven dashed
lines in Fig.~\ref{fig:def5}.  The variables~$\mu_i$ and~$\nu_i$ count
the number of intersections with each line, with~$\nu_i$ corresponding
to lines between the punctures, and~$\mu_i$ lines above and below the
punctures.  (We do not need to know the number of crossings above and
below the first and last punctures, as they can be deduced from the
other crossing information.)

To any loop corresponds a unique set of crossing numbers, and from a
set of valid crossing numbers we can reconstruct a loop.  However, not
all sets of crossing numbers are valid: if we pick some random
non-negative integers for the~$\mu_i$ and~$\nu_i$, then most likely
they will not correspond to a non-self-intersecting closed loop.  To
refine the description, we define \emph{Dynnikov
  coordinates}~\cite{Dynnikov2002} by taking the difference between
adjacent~$\mu_i$ and~$\nu_i$:
\begin{equation*}
  \ac_i = \tfrac12(\mu_{2i}-\mu_{2i-1}),\qquad
  \bc_i = \tfrac12(\nu_{i}-\nu_{i+1}),
\end{equation*}
for~$i = 1,\ldots,\nn-2$.  These coordinates are signed integers; if
we define the vector~$\abv\in\mathbb{Z}^{2\nn-4}$
\begin{equation}
  \abv = (\ac_1,\ldots,\ac_{\nn-2},\bc_1,\ldots,\bc_{\nn-2})
  \label{eq:abvdef}
\end{equation}
then we have a bijection between~$\mathbb{Z}^{2\nn-4}$ and the loops.
This fact is far from obvious: we refer the reader to the literature
for more details~\cite{Dynnikov2002, Moussafir2006, Hall2009,
  Thiffeault2010}, and for the exact prescription of how to
reconstruct a loop from the coordinates.  Figure~\ref{fig:simploop}
shows some examples of the loops corresponding to Dynnikov
coordinates.

The striking fact about the coordinates is that the loop they
represent can be immensely complicated (as tends to happen in chaotic
dynamical systems), and yet only a small number of integers is needed
to represent it.  These integers can, however, be very large.  Thus
the coordinates sidestep one of the biggest computational problems
when dealing with material loops: the memory required to store their
configuration.

\subsection{The action of generators on loops}
\label{sec:action}

\begin{figure}
  \centering
  \includegraphics[width=0.5\figwidth]{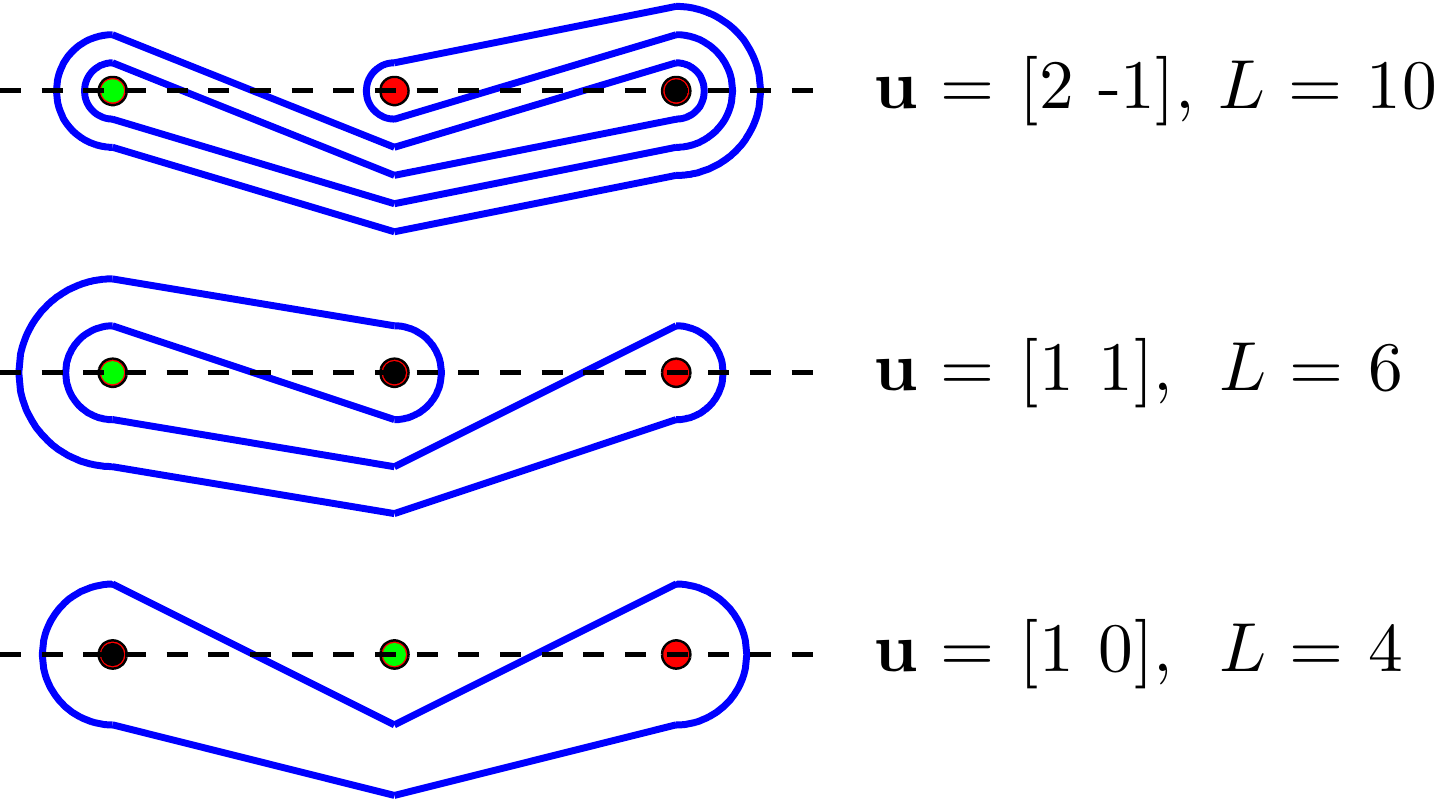}
  \caption{How an initial loop (bottom) around~$\nn=3$ punctures is
    affected by the generators for the trajectories in
    Fig.~\ref{fig:def}. For each loop, the Dynnikov coordinate
    vector~$\abv$ and number of intersections with the dashed
    line~$\Nint(\abv)$ are also given.}
  \label{fig:def3}
\end{figure}

We now come to the final piece of our topological description.  We
have a simple algebraic coding of a set of particle trajectories
(Section~\ref{sec:gen}) and a symbolic representation for topological
loops (Section~\ref{sec:loops}).  In this section, we combine the two
by defining an \emph{action of generators on loops}.
Figure~\ref{fig:def3} illustrates the idea: we start in the bottom
frame with three punctures, surrounded by a topological loop.  Assume
now that the punctures are particle trajectories that move as in
Fig.~\ref{fig:def1}.  The braid representation of their
motion is~$\sigma_1\sigma_2^{-1}$, which means the clockwise
interchange of particles 1 and 2, followed by the counterclockwise
interchange of particles 2 and 3.  This is represented in
Fig.~\ref{fig:def3}, reading from the bottom up to parallel
Fig.~\ref{fig:def4}.  The loop around the punctures is forced to grow
as depicted.

However, we can sidestep having to draw any pictures by using the
\emph{action of generators on loops}.  This action is a set of
algebraic update rules that describes how the Dynnikov coordinates for
a loop are changed by a generator~$\sigma_i$ or~$\sigma_i^{-1}$.
These update rules are somewhat bulky and have been presented
elsewhere~\cite{Dynnikov2002, Moussafir2006, Hall2009}, so we do not
reproduce them here (see also \citet{Thiffeault2010} which follows the
present notation more closely).  The update rules are piecewise-linear
in the coordinates~$(\ac_i,\bc_i)$.

The consequence of these update rules is that given a set of
trajectories we can very rapidly compute the growth of arbitrary loops
enclosing them.  This uses only integer arithmetic.  Contrast this to
evolving material loops: not only does this require a detailed
knowledge of the velocity field, it also requires refinement of the
points making up the exponentially-growing loop, which quickly leads
to memory overflow.  It is this rapid computation of the growth of
loops that allows our detection method to work.

We will associate invariant regions, and thus transport barriers, with
loops having negligible growth.  Hence, we need a way to measure
the length of a topological loop.  \citet{Moussafir2006} demonstrated
that the length of a loop is proportional to the number of times the
loop crosses an imagined line passing through all the punctures (the
line eventually connects to the boundary of the domain).  This line is
shown dashed in Fig.~\ref{fig:def3}; the final number of intersections
is~$10$ (top frame).  The number of intersections~$\Nint(\abv)$ for a
given loop is given by~\cite{Moussafir2006, Hall2009, Thiffeault2010}
\begin{equation}
\Nint(\abv) =
\lvert\ac_1\rvert + \lvert\ac_{\nn-2}\rvert
+ \sum^{n-3}_{i=1} \lvert\ac_{i+1}-\ac_i\rvert + \sum^{\nn-1}_{i=0} \lvert\bc_i\rvert
\label{eq:int}
\end{equation}
where
\begin{align*}
\bc_0 &= -\max_{1\leq i \leq \nn-2} \biggl(\lvert\ac_i\rvert +
  \max(\bc_i,0) + \sum^{i-1}_{j=1}\bc_j\biggr),\\
\bc_{\nn-1} &= -\bc_0 - \sum^{\nn-2}_{i=1}\bc_i\,.
\end{align*}
This formula for `length'~$\Nint(\abv)$ will be used to determine how
fast a loop is growing with time, and thus whether it is a candidate
transport barrier.

Finally, the set of punctures enclosed by a loop that shows negligible
growth, or non-growing loop, will be referred to as an \emph{\IPS}
(IPS).  If the flow inside an invariant region is chaotic, then any
set containing more than two punctures will not be an \IPS.  Hence,
the \IPS\ will automatically be `maximal' and delimit the invariant
region, that is, the only way for a loop not to grow is for it to
contain all the punctures in the region.  \delete{If the flow inside
  an invariant region is not chaotic (for instance, a large vortex),
  then proper subsets of an \IPS\ may also be slow.  In that case we
  can look for the largest \IPS\ (the one containing the most
  punctures).}

\subsection{How the topological tools fit together}
\label{sec:toposumm}

With the tools and definitions developed in the previous sections, it
is now possible to give an overview of the application of braid theory
to the problem of finding two-dimensional transport barriers.  We
start with a set of~$\nn$ two-dimensional particle trajectories on
some domain, obtained from numerical simulations or experimental data.
We then extract a sequence of braid group generators corresponding to
these trajectories.  (This may require some interpolation, but is
usually fairly insensitive to how this is done~\cite{Thiffeault2010}.)

The detection of invariant regions is then reduced to searching
for loops that have negligible growth.  There are two ways of doing
this.  The first method is a methodical search for non-growing
loops by sequentially testing a large number of simple loops, using
their Dynnikov coordinates and the action of generators on these
coordinates; this is presented in Section~\ref{sec:poc}.  This direct
approach is, however, prohibitively slow for more than about eleven
trajectories.  We improve upon it by a refined searching technique,
the pair-loop method, described in Section~\ref{sec:alg}.  We test
both methods using a model system (Section~\ref{sec:sys}).

\section{Search for non-growing loops: Direct method}
\label{sec:poc}

The basis of the direct search method is that if there are loops that
have negligible growth, which correspond to transport barriers, they
will emerge by examining the set of all simple loops.  Here we define
what we mean by a `simple' loop, give an outline of the algorithm to
analyze the loops, and apply this to a model system.

\subsection{Simple loops}

\begin{figure}
  \centering
  {\includegraphics[width=0.7\figwidth]{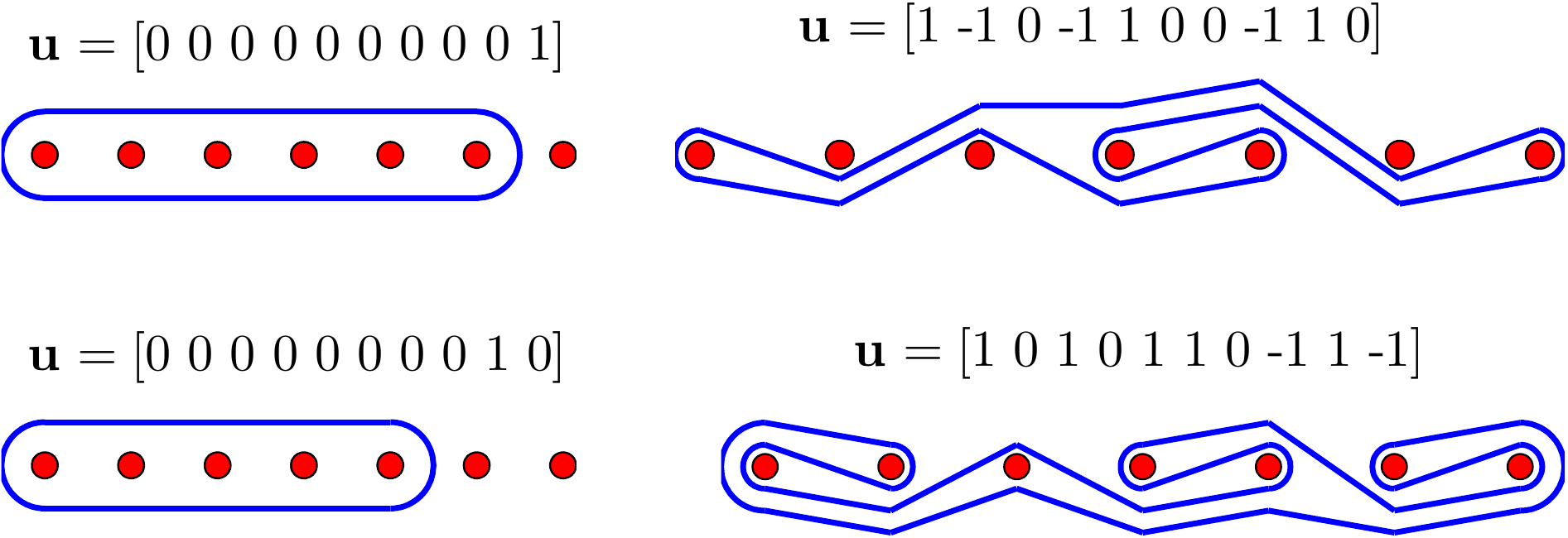}}    
  \caption{Four simple loops with Dynnikov coordinates between~$-1$
    and~$1$.}
  \label{fig:simploop}
\end{figure}

We assume that the invariant regions present in our flow are described
by relatively simple loops.  All loops can be described by their
Dynnikov coordinates, and the complexity of a loop is typically
reflected in the magnitude of the coordinates.  In fact, as a loop
becomes more convoluted its coordinate values grow larger.  For this
reason, and to make the problem computationally tractable, we limit
our search to loops with initial values of the Dynnikov coordinates
between~$-1$ and~$1$.  Four such loops are shown in
Fig.~\ref{fig:simploop}, demonstrating the range in complexity
captured by this limited set of Dynnikov coordinates.  As mentioned in
Section~\ref{sec:loops}, a set of~$\nn$ trajectories has loops
represented by $2\nn-4$ coordinates.  This results in $3^{2\nn-4}-1$
possible loops with coordinates between~$-1$ and~$1$.  (The null loop
with all coordinates equal to zero is not considered.)  Many of these
loops are `multi-loops' (see Fig.~\ref{fig:structures}c) which are
redundant, but it is not easy to eliminate those \emph{a priori} from
the Dynnikov coordinates, so we test those as well.

\subsection{Algorithm outline}
\label{sec:algout}

In order to find the non-growing loops, we developed an algorithm in
Matlab.  The objective of this algorithm is to take in a set of
trajectories and output the non-growing loops surrounding
punctures.  The first step is to calculate the generator sequence
which defines the braid of the trajectories.  We discuss this in
Section~\ref{sec:gen}, and we refer the reader to
\citet{Thiffeault2005, Thiffeault2010} for more details and Matlab
codes.

Next, for each Dynnikov coordinate vector with entries between~$-1$
and~$1$ inclusively, we apply the entire generator sequence to each
loop and compute the intersection number~$\Nint(\abv)$, which is
proportional to the loop's length
(Section~\ref{sec:loops}--\ref{sec:action}).  We then identify the
loops with small final number of intersections as non-growing
loops.  As we shall see below, `small' typically means tens of
intersections, compared to most loops which grow to have~$10^{10}$ or
more intersections.  The punctures enclosed by a non-growing loop
form an \IPS\ (see Section~\ref{sec:action}).  

\subsection{Modified Duffing oscillator}
\label{sec:sys} 

In order to test the direct method, we use a model velocity field with
transport barriers.  This velocity field will be used to calculate
trajectories for analysis.  While this is not the target application
of the algorithm, it is useful for testing to have a system where we
can determine {\it a priori} the location of transport barriers.

For this test, we use a slightly modified form of the Duffing
oscillator.  One change is made in order to create two
relatively-complex invariant regions, and the other change adds solid
body rotation to further obscure the dynamics.  The basic system
without rotation is defined as
\begin{subequations}
\begin{align}
\dot{x} &= y+\alpha \cos\omega t, \\
\dot{y} &= x(1-x^2)+\gamma \cos\omega t - \delta y\,,
\end{align}
\label{eq:Duffing}%
\end{subequations}
where we will use $\alpha=.1$, $\gamma = .14$, $\delta = .08$, and
$\omega=1$.  To add rotation to the system, and thus further `hide'
the transport barriers, we make the transformation
\begin{equation*}
\begin{pmatrix}\tilde x\\ \tilde y\end{pmatrix}
= \begin{pmatrix}
 \cos\Omega t & \sin\Omega t \\
-\sin\Omega t & \cos\Omega t
\end{pmatrix}
\begin{pmatrix} x\\  y\end{pmatrix}.
\end{equation*}
Note that with~$\delta\ne0$ this velocity field is \emph{not}
incompressible.  From a topological viewpoint, this is irrelevant to
the issue of invariant regions, and shows the wide applicability of
the method.  More importantly, the transport barriers have a more
complex shape with these parameter values.  A more realistic,
incompressible flow will be used in Section~\ref{sec:rod}.

\begin{figure}
  \centering
  {\includegraphics[width=0.45\figwidth]{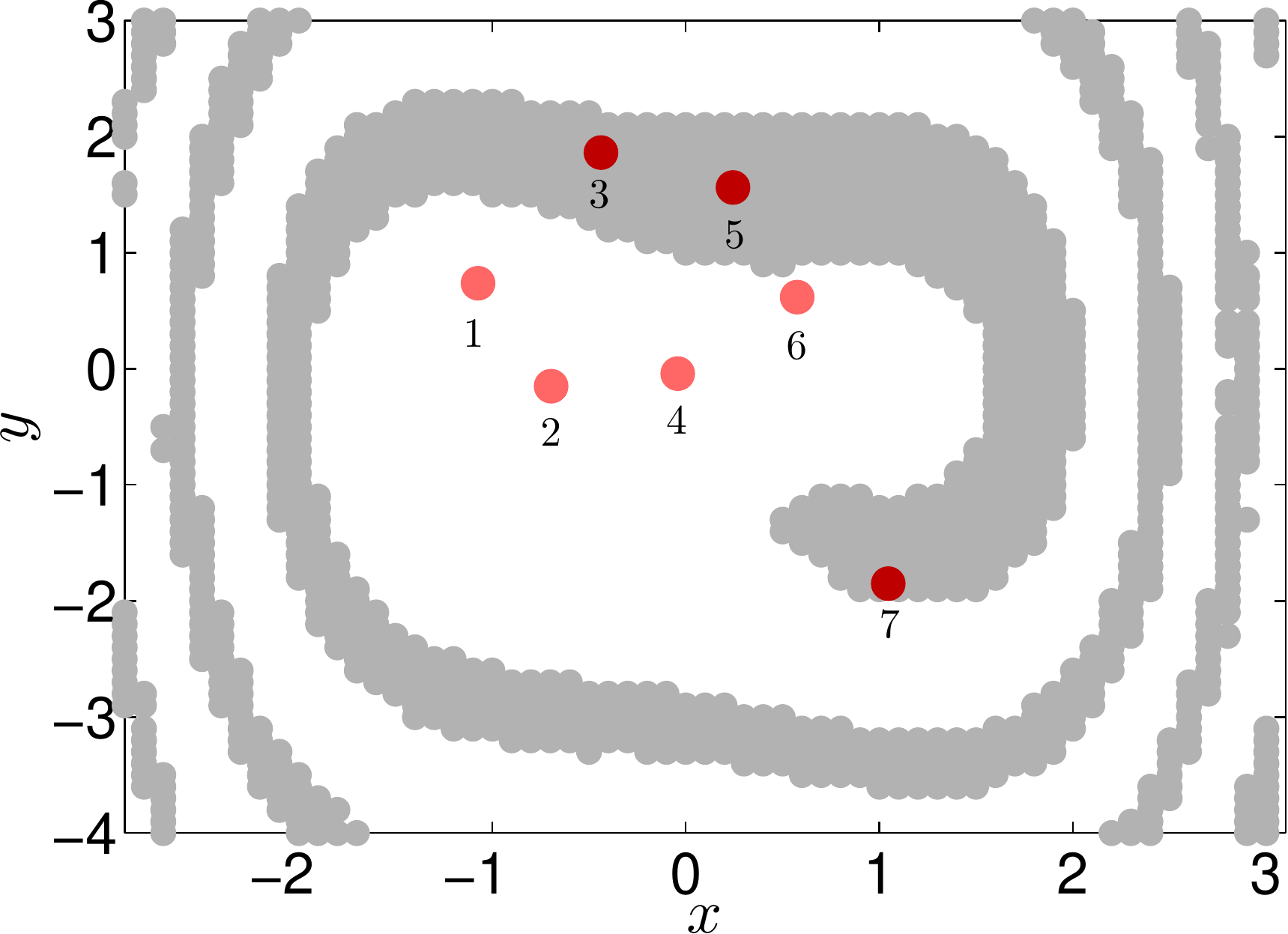}}
  \caption{The two types of initial conditions for the modified
    Duffing oscillator system~\eqref{eq:Duffing} with rotation.
    Orbits in the white region are attracted to a chaotic
    region at the center.  Orbits in the gray region approach a limit
    cycle around the white region.  The dots (numbered from left to
    right) are the initial conditions for the trajectories discussed
    in Section~\ref{sec:pocres}. }
  \label{fig:regions}
\end{figure}

The key feature of the Duffing system with these parameter values is
that there are two separate regions, plotted in Fig.~\ref{fig:regions}
at~$t=0$.  There is a transport barrier between the gray and the white
regions, that is, all points initially in the gray region stay in
their own region, and vice-versa.  Thus, a set of trajectories
entirely within one region should have a non-growing loop
surrounding them.

\subsection{Results of the direct method}
\label{sec:pocres}

To test the direct method we used a variety of sets of initial
conditions.  A single representative example is presented here.  This
system consists of seven trajectories with initial conditions shown in
Fig.~\ref{fig:regions}.  Based on the initial conditions, four of the
trajectories are contained in the white region (chaotic), and the
other three trajectories are in the gray region (limit cycle).  From
these trajectories, we compute the corresponding braid group
generators (Section~\ref{sec:gen}) and examine the growth of $3^{10}$
loops via the update rules (Section~\ref{sec:action}).

\begin{figure}
  \centering
  \subfloat[]{\label{fig:bres1}\includegraphics[width=.3\textheight]{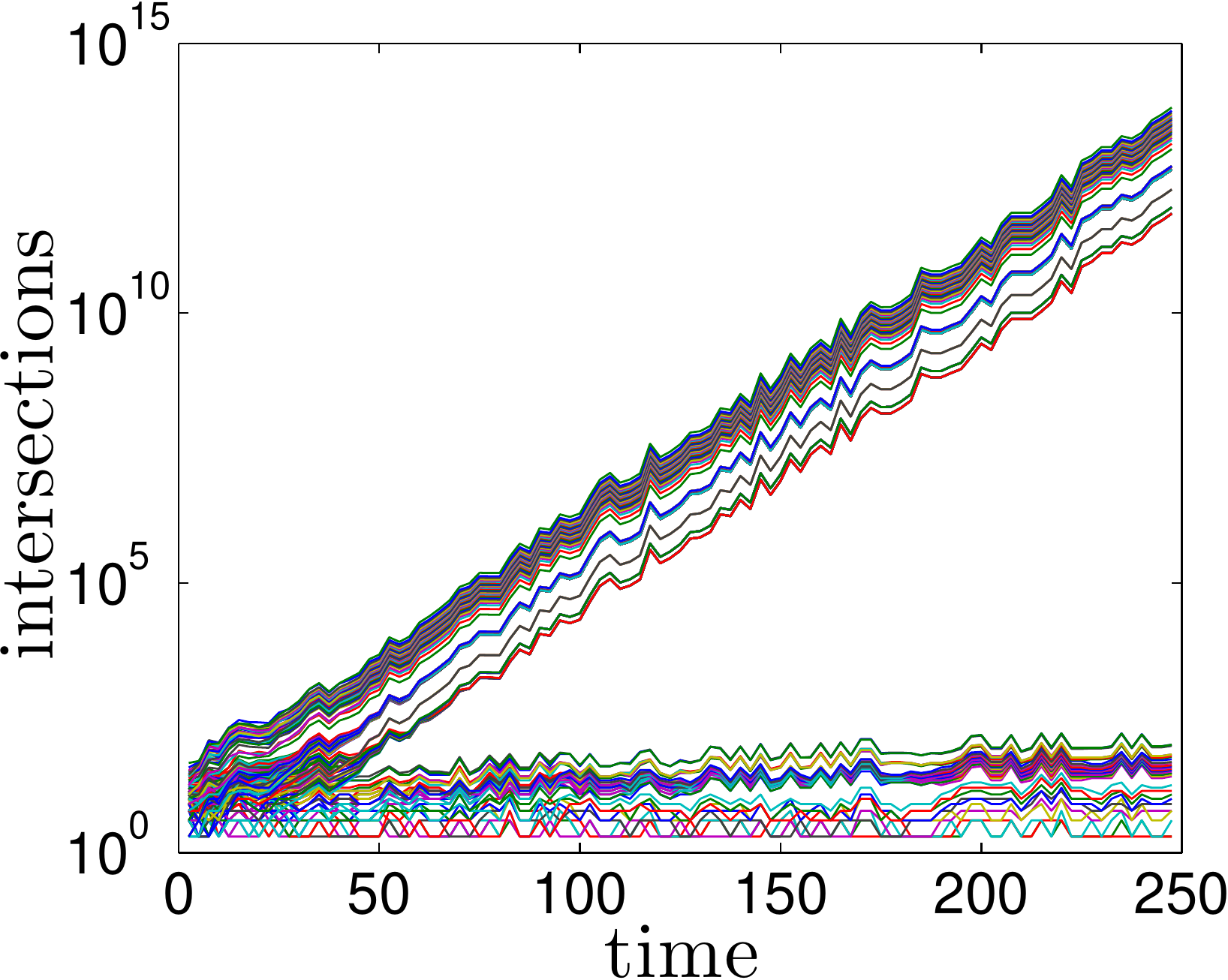}}
  \hspace{.03\figwidth}
  \subfloat[]{\label{fig:bres2}\includegraphics[width=.285\textheight]{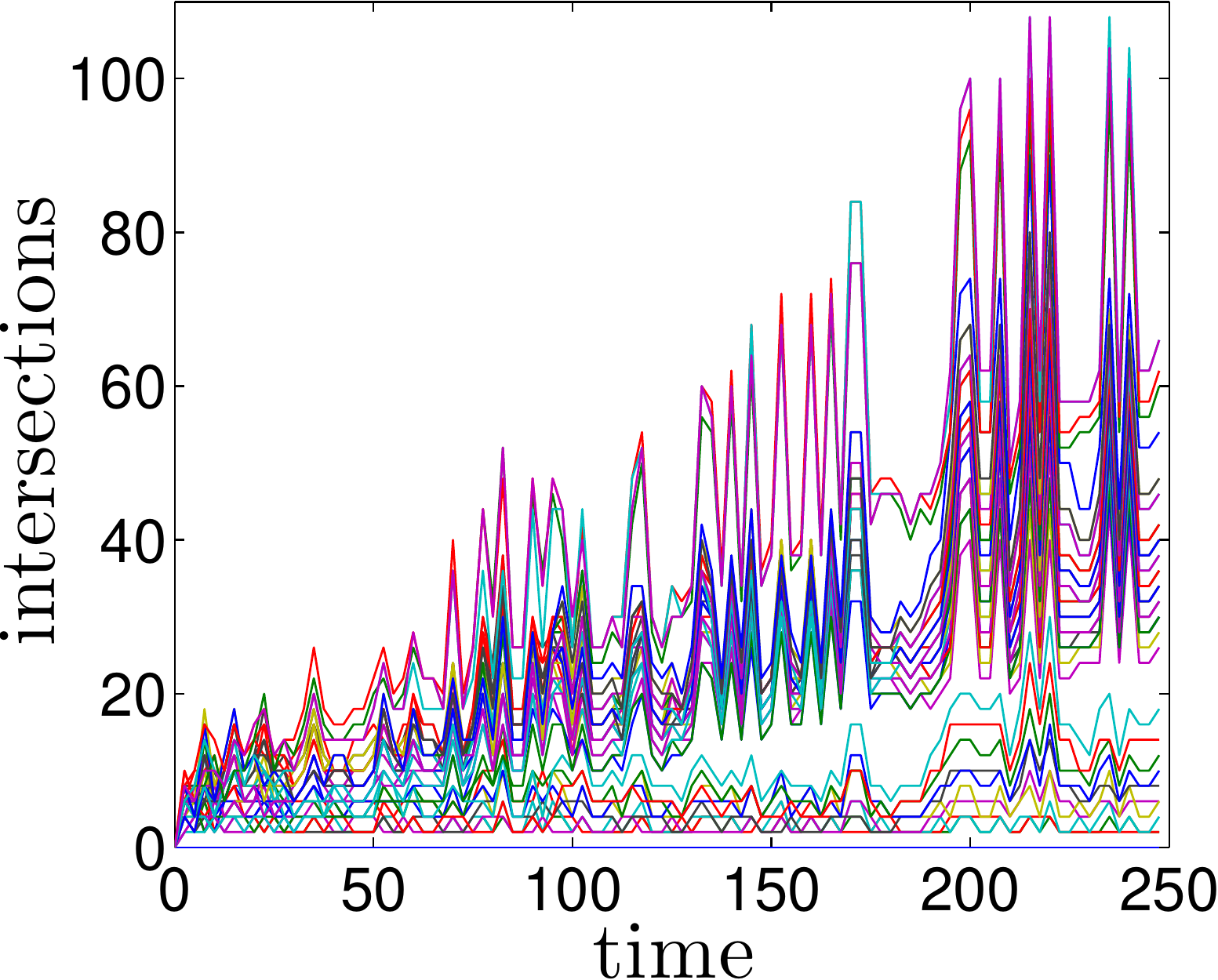}}
  \caption{(a) Intersections with the~$x$-axis as a function of time
    for loops with Dynnikov coordinate values ranging from~$-1$
    to~$1$. (b) Zoom on a linear scale of the loops in (a) that have
    sub-exponential growth.}
  \label{fig:bres}
\end{figure}

We plot the number of intersections with the~$x$-axis as a function of
time for a representative set of loops in Fig.~\ref{fig:bres1}.
(Recall from Section~\ref{sec:loops} that these intersections are a
proxy for length.) This plot indicates that there are two regimes of
growth: loops growing exponentially (linearly increasing in the
semi-log plot) and loops that show sub-exponential growth
(approximately flat in the plot).  The loops growing exponentially do
not correspond to loops surrounding an invariant region, so they can
be discarded.  It should be noted that all the discarded loops enclose
a proper subset of the four trajectories in the central chaotic region
and a subset of the three limit-cycle trajectories.

\begin{figure}
\centering
\includegraphics[width=0.45\figwidth]{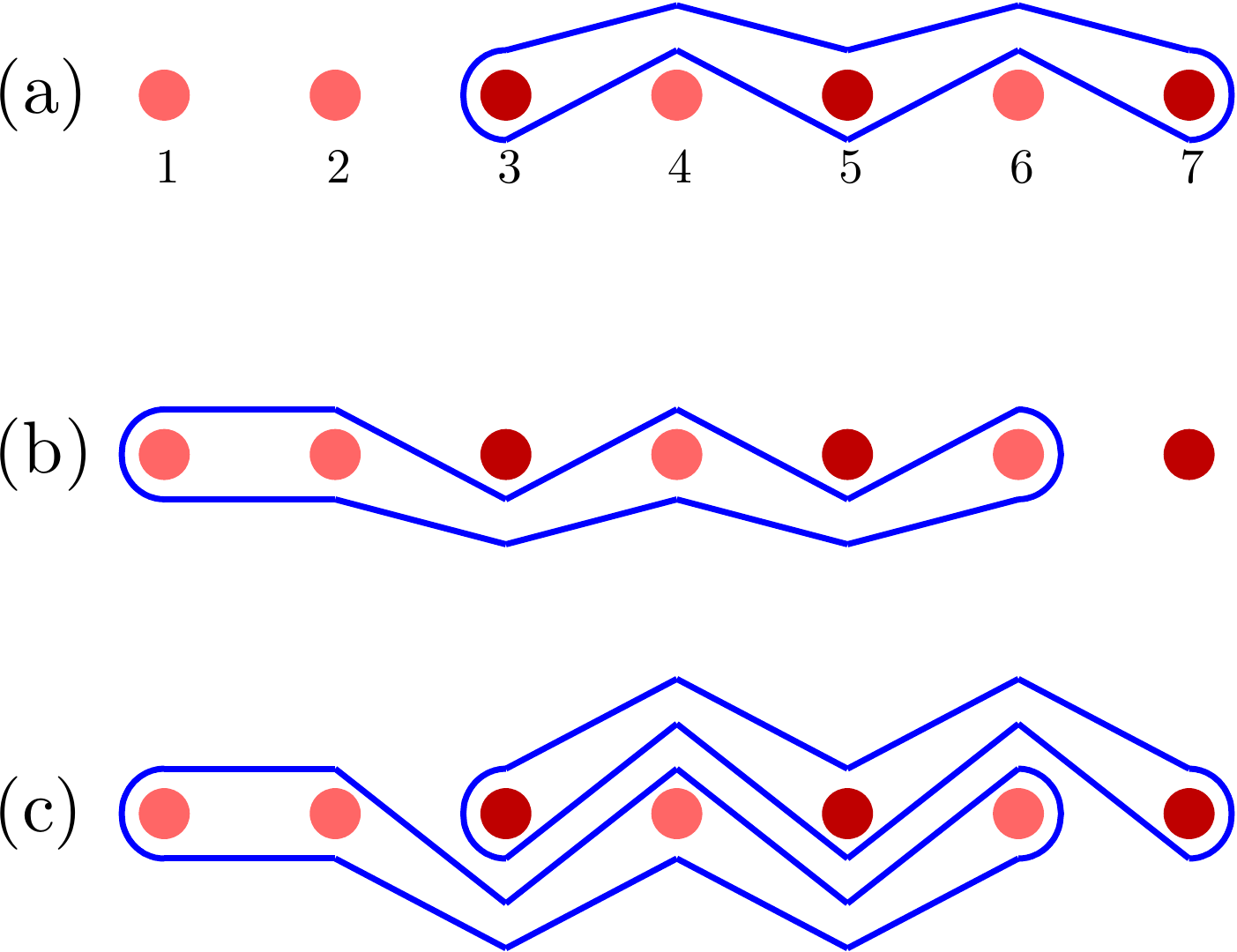}
\caption{Simplified loop diagrams of the loops which do not show
  steady growth.  (a) Loop containing the limit-cycle trajectories
  (\protect\coloronline{red}{dark} punctures --- 3, 5, and 7).  (b) Loop
  containing the inner chaotic trajectories
  (\protect\coloronline{pink}{light} punctures --- 1, 2, 4, and 6).  (c)
  Double-loop consisting of the (a) and (b) loops.}
\label{fig:structures}
\end{figure}

A closer view for the loops with non-exponentially-growing
intersections is shown in Fig.~\ref{fig:bres2}.  This plot is on a
linear scale and shows that some of the loops grow approximately
linearly while others don't grow at all.  The slowly-growing loops
that grow linearly contain all four trajectories in the chaotic region
and a proper subset of trajectories from the region with a limit
cycle.  The non-growing loops, shown in Fig.~\ref{fig:structures},
correspond to transport barriers.  Changes in the length of the loops
do occur due to motion of the particles, but there is no steady growth
of the initial loop.

Thus, the direct method found the expected `topological transport
boundaries' by sifting through an enormous number of loops, and
finding the very few that do not grow.  But a great drawback of this
method is how poorly it scales as the number of trajectories is
increased: the number of loops examined is $3^{2\nn-4}\sim 9^\nn$.
For this algorithm, the bottleneck in the calculation is the time
required to apply the generators to all the loops, which scales with
the number of loops.  The direct method is rendered useless when
dealing with more than eleven trajectories or so, which leads us to
improve the detection method in Section~\ref{sec:alg}.

\section{Search for non-growing loops: Pair-loop method}
\label{sec:alg}

While the methodical search through simple loops produces the
non-growing and slowly-growing loops, it is not practical for more
than a dozen punctures.  In real applications we might have access to
hundreds of trajectories, so we need the capacity to easily analyze
many more than a dozen trajectories.

One major drawback of the previous method is that no information is
gained from a loop other than its growth rate, so every loop has to be
tested.  The refined algorithm tests a specific set of loops and
analyzes their state after the generators have been applied.
Specifically, the loops analyzed are the ones that connect two
punctures, which we call \emph{pair-loops}.  Their final state gives
an indication of which punctures cause the loop to grow and which
punctures become \emph{entangled} by a given loop (we will define
entanglement more precisely in Section~\ref{sec:step2}).  This is the
basis for the pair-loop algorithm.

The objective of the pair-loop method is the same as the direct method
of Section~\ref{sec:poc}: to take a set of trajectories and find
non-growing loops.  The steps of the pair-loop method are as follows:
\begin{enumerate}

\item We calculate the generator sequence corresponding to the
  trajectories.  We then apply this sequence to pair-loops via their
  Dynnikov coordinates (Section~\ref{sec:step1}).

\item We then analyze the resulting final loops to determine which
  punctures are entangled by each loop, yielding a collection of
  \EPSs\ (Section~\ref{sec:step2}).

\item With the information on \EPSs, we determine which punctures 
are contained in invariant regions (Section~\ref{sec:step3}).

\item We construct the loop surrounding each invariant
  region (Section~\ref{sec:step4}).

\end{enumerate}
After describing the steps of the improved algorithm in
Section~\ref{sec:step1}--\ref{sec:step4}, we apply it to the
modified Duffing oscillator in Section~\ref{sec:DuffApp}.

\subsection{Step 1: Application of generators to pair-loops}
\label{sec:step1}

\begin{figure*}
  \centering
  \includegraphics[width=0.75\figwidth]{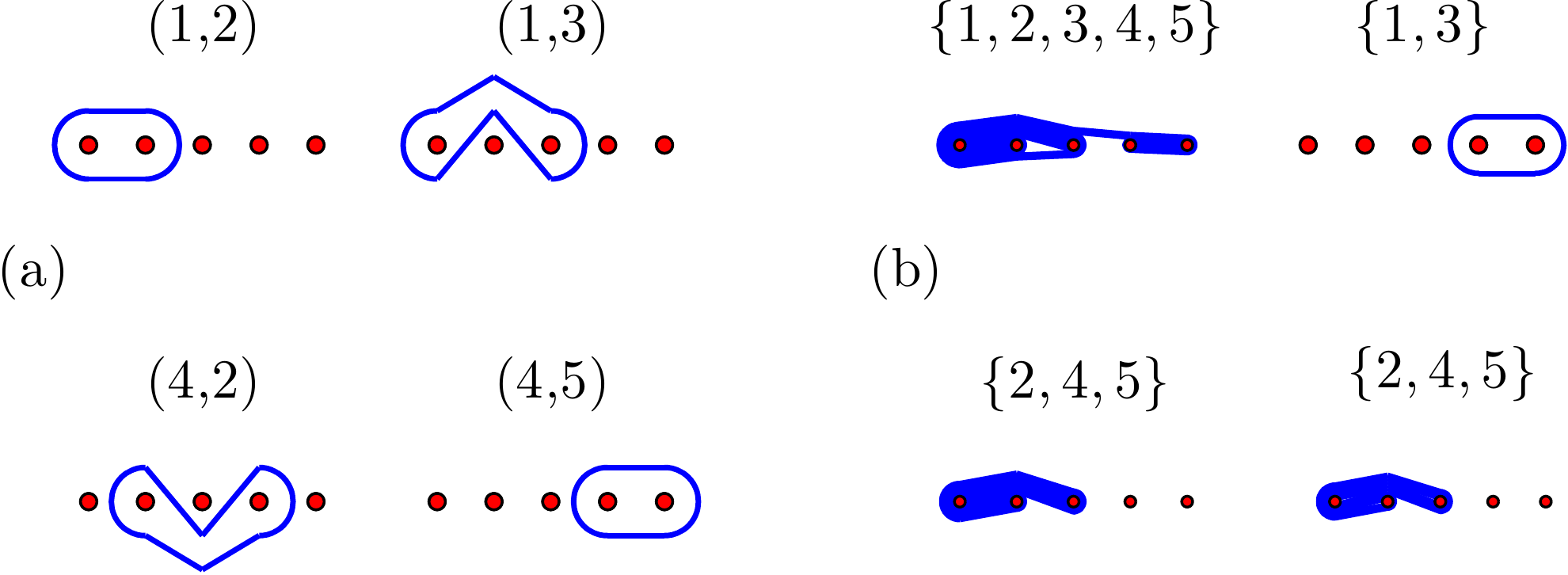}
  \caption{(a) A representative set of the initial pair-loops for a
    five-puncture system.  (b) The loops that result after applying a
    sequence of~$143$ generators, and the corresponding \EPS.  The
    loops (1,3), (4,2), and (4,5) fail to entangle all punctures after
    applying the generators.  This group of trajectories form the two
    \IPSs~$\{1,3\}$ and~$\{2,4,5\}$.}
  \label{fig:dis}
\end{figure*}

As mentioned in the description of the algorithm, in the pair-loop
method we start with loops that connect pairs of punctures.  While
there are an infinite number of ways for a loop to connect punctures,
the main concern here is how simple loops entangle the other
punctures.  Given this consideration, punctures are connected as shown
in Fig.~\ref{fig:dis}a.  Adjacent punctures are connected as for
loop~$(1,2)$ in that figure.  For punctures that are not adjacent, two
loops are created where one loop passes above intermediate punctures
(e.g. loop~$(1,3)$ in Fig.~\ref{fig:dis}a), and the other passes below
(loop~$(4,2)$ in the figure).  The notation for each initial pair-loop
is $(p,q)$, where if $p<q$ (resp., $p>q$) a loop connects puncture~$p$
to~$q$ and passes above (resp., below) intermediate punctures.

\begin{figure}
  \centering
  \includegraphics[width=0.7\figwidth]{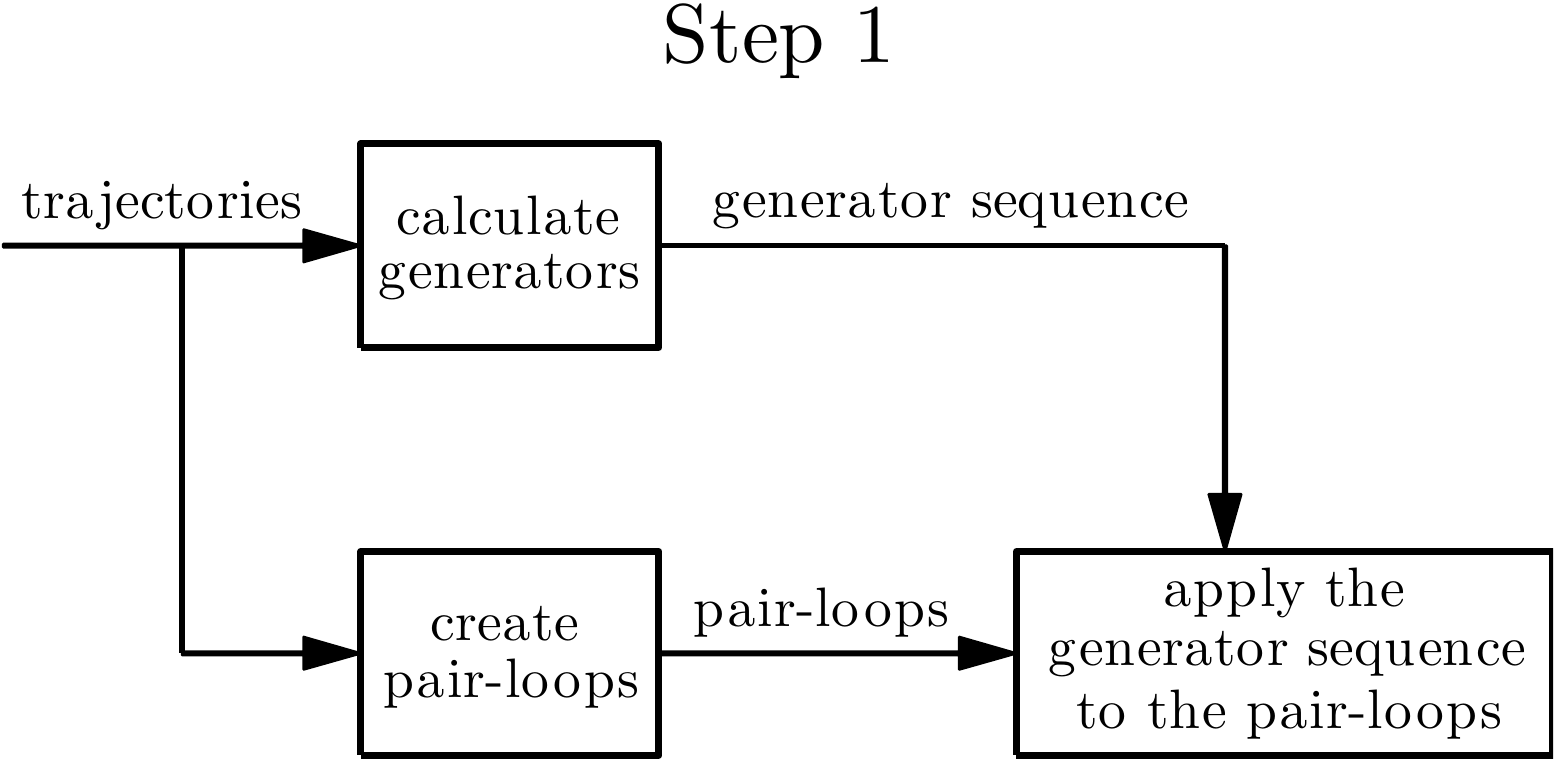}
  \caption{Step 1 takes the trajectories as the input and outputs a
    set of Dynnikov coordinates representing the pair-loops after the
    application of the generator sequence.}
  \label{fig:bd1}
\end{figure}

The first step of the method, shown diagrammatically in
Fig.~\ref{fig:bd1}, takes the trajectories as input.  As in the direct
method, the generator sequence is calculated from the trajectories.
Then we determine the pair-loops to be tested.  For a system of $\nn$
trajectories, there are $\nn-1$ loops connecting adjacent punctures
and $(\nn-2)(\nn-1)$ loops connecting non-adjacent punctures above and
below, for a total of $(\nn-1)^2$ loops to analyze.  The generator
sequence is then applied to the pair-loops and the resulting Dynnikov
coordinates are the output of this step. A sample set of pair-loops
before and after the application of the generator sequence is shown
in Fig.~\ref{fig:dis}.

At this point, a marked improvement has already been made.  In the
direct method, the bottleneck was the application of the generators to
$3^{2\nn-4}-1$ loops, causing the run time to scale exponentially
as~$t\sim O(9^\nn)$.  With the pair-loop approach, the bottleneck of
the algorithm still involves the number of loops analyzed, but the
number of loops tested now scales algebraically as $O(\nn^2)$.

\subsection{Step 2: Determine the \EPSs}
\label{sec:step2}

\begin{figure}
  \centering
  \includegraphics[width=0.7\figwidth]{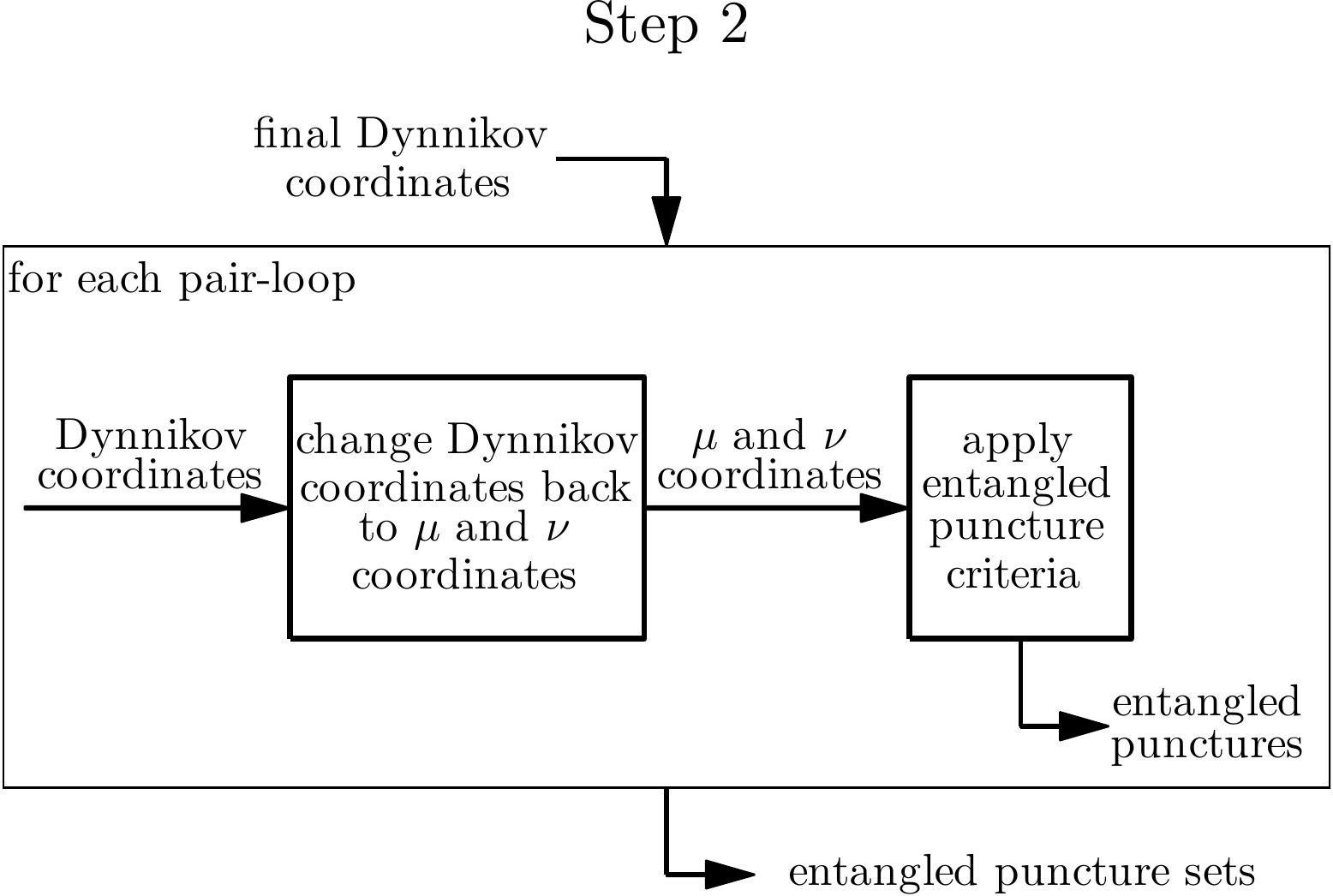}
  \caption{Step 2 takes the pair-loops and finds the \EPSs.}
  \label{fig:bd2}
\end{figure}

After the generator sequence has been applied to the pair-loops, we
need to determine which punctures are entangled by the loops.  In this
step, presented in Fig.~\ref{fig:bd2}, we go through each set of
Dynnikov coordinates, change to crossing number coordinates, apply the
entangled puncture criterion, and determine which punctures are
entangled.  We say that a puncture is \emph{entangled} by a loop if
the loop passes both above and below the puncture, with punctures
ordered from left to right as usual.  To determine this, the Dynnikov
coordinates of a loop are converted back to the $\mu,\nu$
representation (see~\citet{Hall2009} for the inversion formula).  The
odd-indexed crossing numbers $\mu_{2i-1}$ count the number of times
the loop passes above a puncture, and the even-indexed crossing
numbers $\mu_{2i}$ count the number of times the loop passes below the
puncture (see Fig.~\ref{fig:loop}).

With this $\mu,\nu$ coordinate representation, it is straightforward
to determine if a puncture is entangled or not: the $i$th puncture is
entangled if both $\mu_{2i-1}$ and $\mu_{2i}$ are non-zero for the
loop.  For each pair-loop, the set of all entangled punctures is the
\emph{\EPS} (EPS) of that loop.  After finding the \EPS\ for all loops, we
remove duplicates and only consider \ESs\ which are a proper subset of
the full set of punctures.  (That is, we discard \EPSs\ that contain
all the punctures.)

For the example presented in Fig.~\ref{fig:dis}, the generators
applied to the loops in Fig.~\ref{fig:dis}a produce the loops in
Fig.~\ref{fig:dis}b.  Of the four loops shown, one entangles all
punctures, so this \EPS\ does not provide any information about the
invariant regions and is discarded.  The loop~$(1,3)$ entangles only
punctures~$1$ and~$3$, so its \EPS\ is~$\{1,3\}$.  The loops~$(4,2)$
and~$(4,5)$ entangle only the punctures~$2$, $4$, and~$5$, so their
\EPS\ is~$\{2,4,5\}$.  The example results in two \ESs\ which also
happen to be the punctures in the two invariant regions for this
system.  There is a non-growing loop around punctures~$1$ and~$3$ and
one around~$2$, $4$, and $5$.  Thus, in this case \EPSs\ are also
\IPSs.

\subsection{Step 3: Find the \IPSs }
\label{sec:step3}

\begin{figure}
  \centering
  \includegraphics[width=0.7\figwidth]{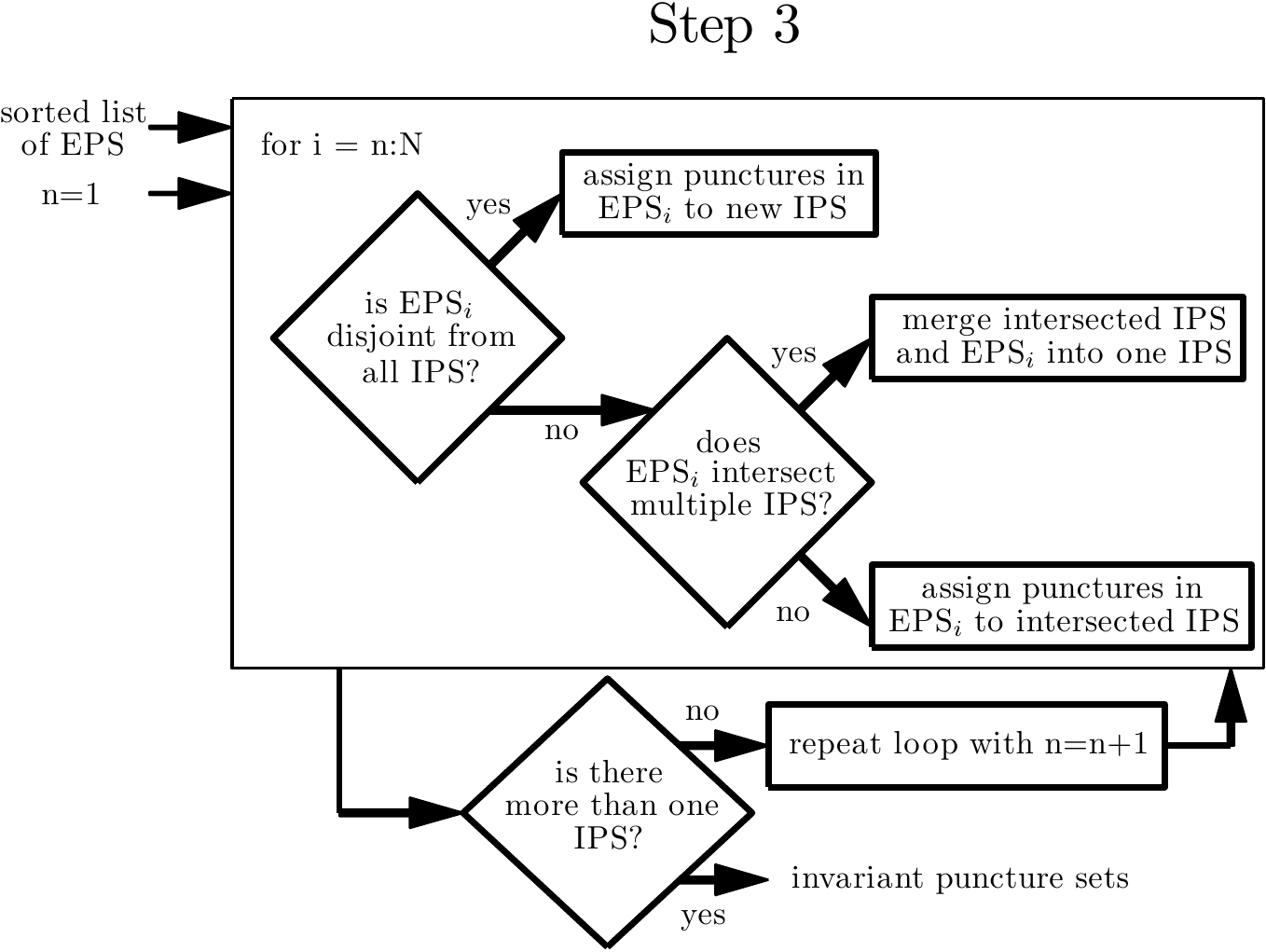}
  \caption{Step 3 takes the \EPSs\ and distributes them into the
    \IPSs.}
  \label{fig:bd3}
\end{figure}

While the example presented in Sections \ref{sec:step1} and
\ref{sec:step2} produces the \IPSs\ directly from the \EPSs, this is
not always the case.  For a pair-loop, the two punctures are either
within the same invariant region or not.  If a pair-loop connects
points within an invariant region, it can either entangle all the
punctures within the region, a subset of those punctures, or a subset
as well as punctures from another invariant region.  In the simple
example, loops connecting punctures within a region always only
entangled with all the punctures within the same region.
Alternatively, if a loop connects punctures from different invariant
regions, they will entangle punctures from at least those two regions.
This step of the method, presented in Fig.~\ref{fig:bd3}, takes in the
\EPSs\ and deciphers which punctures form an invariant region.

To create the \IPSs, the code runs through the \ESs\ and allocates a
region index to each puncture in the system.  The \ESs\ are read in
from largest to smallest (in terms of the number of punctures they
contain), so the first and largest set (if there are multiple largest
sets the starting set is chosen arbitrarily) has all its punctures
designated as region~$1$, and then the next \ES\ is analyzed.  If the
second set intersects the first, then all points in the second set are
labeled with the same region index.  Otherwise, the second set is
disjoint from the first region, and it is given a new region index.
This process is repeated for each of the remaining \EPSs, merging the
sets if they share points, or otherwise creating a new region index.
It is possible for an \ES\ to intersect multiple regions, in which
case the intersected regions as well as the punctures in the \ES\ are
all merged into one region.

At the end of this process, we obtain a list of disjoint puncture
sets.  If this list has only one member (containing all the
punctures), then we have failed to find any invariant regions, and we
restart the process of finding the \IPSs.  But now instead of starting
with the largest \ES, we start with the next largest \ES\ as our first
region, completely discounting the previous largest one.  We repeat
until we obtain a list of several disjoint \IPSs.  If we fail to
obtain such a list even after using all the \ESs\ as a starting point,
then we conclude that there are no non-growing loops present.

This method of comparing \ESs\ to discern invariant regions is
susceptible to the unintentional merging of multiple regions into one
larger region.  To combat this issue, the entire pair-loop method can
be run on each \IPS\ produced, to determine if it can be broken down
into smaller invariant regions.

For clarity, the following is an example where steps 1 and 2 have
produced the following \ESs:
\begin{equation}
  \{1,2,3,4\}, \quad
  \{3,4,5,6\}, \quad
  \{1,2\}, \quad
  \{3,4\}, \quad
  \{5,6\}.
  \label{eq:EPSex}
\end{equation}
The search for the \IPSs\ begins by assigning the punctures in the
first \ES\ the region index 1.  Next we determine that the second \ES,
$\{3,4,5,6\}$, intersects region 1, so all of the punctures in the
second \ES\ are also assigned the region index 1.  This means that all
punctures are in the same region, indicating that the search failed
and we must start again with the next largest set.  This time we start
with the \ES\ $\{3,4,5,6\}$, assign it the region index 1, and move on
to the \ES\ $\{1,2\}$.  This set is disjoint from all the assigned
regions, so it is given the region index 2.  The code then compares
the two remaining sets, but both of these are subsets of region 1.
This time the code successfully finds multiple disjoint \IPSs, and
moves on to the next step of the algorithm.  It may be the case that
$\{3,4\}$ and $\{5,6\}$ are actually disjoint \IPSs; this is
discovered by running the entire pair-loop method solely on the
$\{3,4,5,6\}$ trajectories.

\subsection{Step 4: Creation of slow-growing loops}
\label{sec:step4}

At this point in the procedure, we have successfully identified the
punctures that make up the \IPSs.  Because the regions are assumed to
be relatively simple, it is reasonable to presume that the
least-complicated loop enclosing the punctures in the \IPSs\ will be
the non-growing loop.  However, there are an infinite number of ways
to connect the punctures, but only one way in which the loop will not
grow.  This step determines the loop by creating link-loops, which
will be described next.  Because of its complexity, this step is only
necessary if the ``obvious'' loop surrounding the region turns out to
be a growing loop.  For this reason, many readers may wish to skip the
rest of this section.

\begin{figure}
  \centering
  \includegraphics[width=0.35\figwidth]{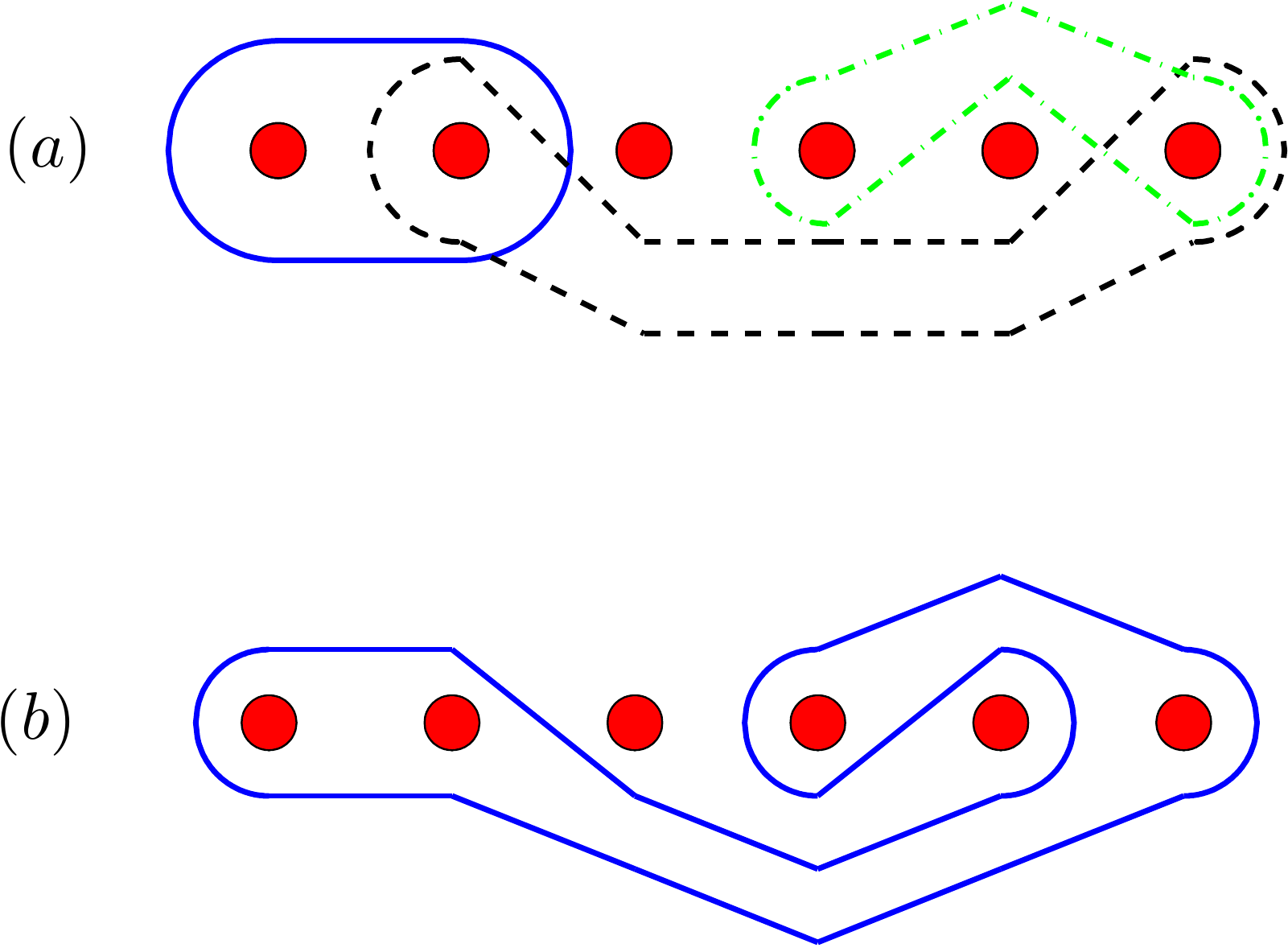}
  \caption{Punctures $1$, $2$, $4$, and $6$ are in an invariant
    region.  (a) The link-loops shown here connect the punctures.
    The solid line connects $1$ to $2$, dashed connects $2$ to $6$,
    and dot-dashed connects $6$ to $4$. (b) The connecting
    link-loops can be merged to form a single loop around the
    punctures.}
  \label{fig:dync}
\end{figure}

A loop enclosing the \IPS\ can be viewed as the union of the
pair-loops connecting punctures in a region.  These specific
pair-loops are referred to as \emph{link-loops}.  An example is
presented in Fig.~\ref{fig:dync}, where three link-loops connect the
punctures in a region, then are merged to form one loop.  These
link-loops are the key to finding the non-growing loop surrounding the
entire region.

A key characteristic of the non-growing loop is that punctures from
outside the region do not cause the loop to grow, nor do they
  cause the link-loops to grow.  However, the other punctures in the
\IPS, but outside a given link-loop, will cause that link-loop to
grow.  This property is the basis for the procedure to find the
correct link-loops and to build the non-growing loop.

\begin{figure}
  \centering
  \includegraphics[width=0.7\figwidth]{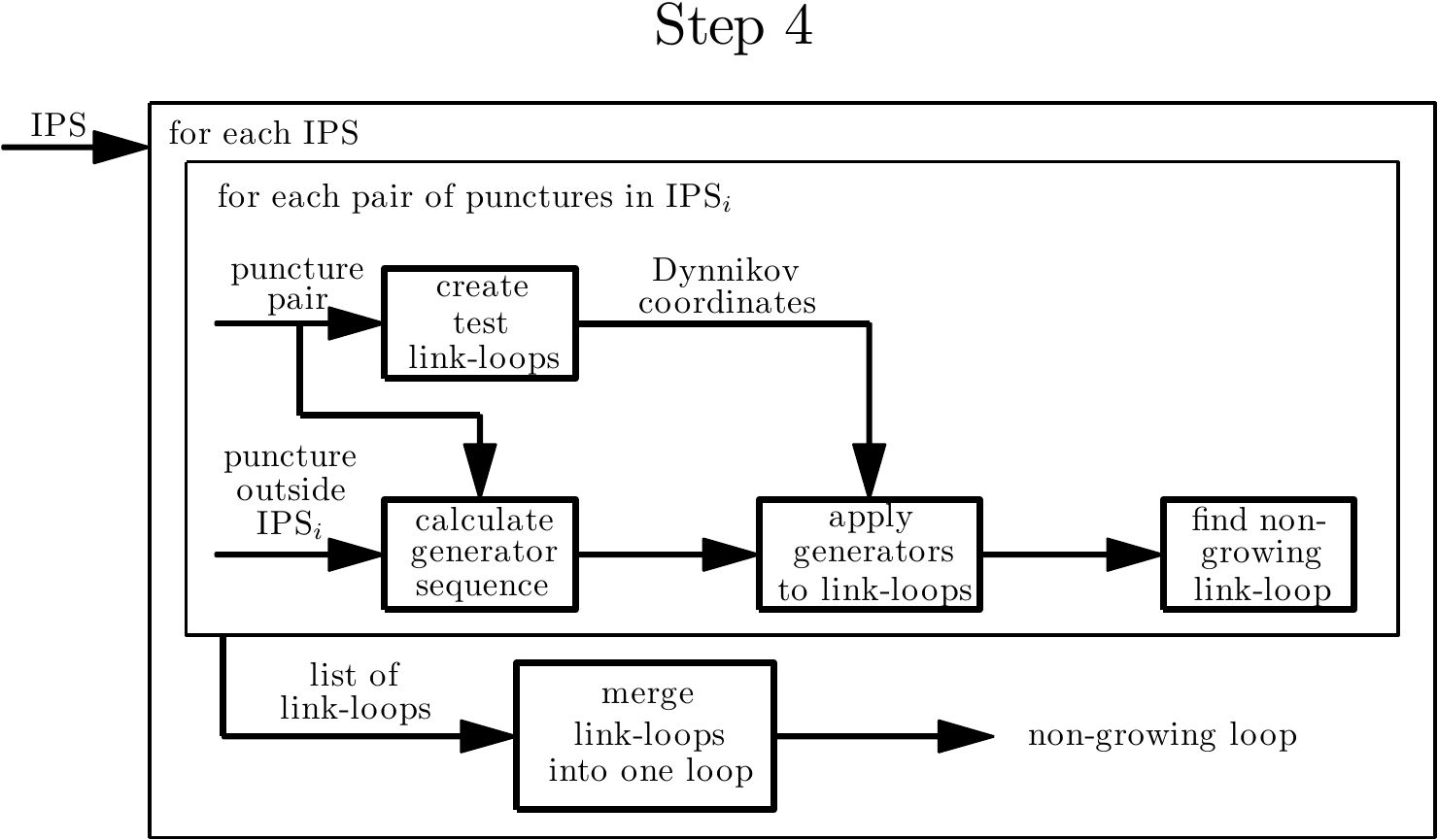}
  \caption{Step 4 takes the \IPSs\ and finds the corresponding
    non-growing loops.}
  \label{fig:bd4}
\end{figure}

This step takes as input the \IPSs\ and finds the link-loops
connecting punctures within the region.  This search is done by
considering the topological sub-braid resulting from the two selected
punctures within the region and all punctures outside of the region.
If a link-loop is present between the two punctures, there will be a
loop connecting the two punctures that does not grow.  Once link-loops
are found connecting all the punctures in the \IPSs, they are combined
to form the non-growing loop.  This step is outlined in
Fig.~\ref{fig:bd4}.

\subsection{Application to the modified Duffing system}
\label{sec:DuffApp}

The pair-loop method is much faster at finding non-growing loops,
making it practical to use braid theory to detect transport barriers.
The run times for the direct method (Table~\ref{tab:times})
significantly limit its applicability, but by improving the scaling
from $O(9^\nn)$ to $O(n^2)$ the pair-loop method is a viable
alternative.  Table~\ref{tab:times} compares the run times of both
methods when applied to the modified Duffing oscillator of
Section~\ref{sec:sys}, for increasing number of punctures~$\nn$.
Because of the extra overhead of the pair-loop algorithm, it is slower
until about~$9$ punctures, after which it scales much better that the
direct method.

\begin{table*}
\centering
\begin{tabular}{lllllllll}
  \hline
  \# of trajectories & 5 & 6 & 7 & 8 & 9 & 10 & 11 & 20 \\
  \hline
  direct method \qquad{} & 0.33 \quad{} & 0.46 \quad{} & 0.70 \quad{}
  & 6.0 \quad{} & 53 \quad{} & 462 \quad{} & 3445 \quad{} &
  N/A \\
  pair-loop method & 6.7 \quad & 9.5 \quad & 11.6 \quad & 12.3 \quad &
  13 \quad & 15 \quad & 20 \quad & 128 \\
  \hline
\end{tabular}
\caption{Run times (in seconds) for the modified Duffing oscillator using
  the direct and pair-loop methods.  The run with 20 trajectories had
  over four times as many generators than previous cases, which
  contributes to the run time being longer than the trend would suggest.}
\label{tab:times}
\end{table*}

While the run times for the direct method clearly scale with the
predicted $t\sim O(9^\nn)$, the pair-loop algorithm does not
demonstrate a specific trend in growth time.  This is due to the fact
that we are not yet in the regime where the number of loops analyzed
is the limiting factor.  For systems where there $\nn>50$, the number
of pair-loops analyzed is the bottleneck.

\section{Finding non-growing loops in a rod-stirring device}
\label{sec:rod}

We finally apply the method to an actual fluid system.  We use the
translating-rotating mixer (TRM) system described
by~\citet{MattFinn2001}.  Their two-dimensional stirring mechanism
features a circular domain of viscous fluid and a circular stirring
rod, which can translate and rotate about its center.  The authors
derive an analytic complex-variable solution for the instantaneous
velocity field of the fluid in the zero-Reynolds-number limit, which
allows an accurate calculation of fluid trajectories.

For many parameter values, islands of poor mixing are present for
periodic rod motions.  Each island is surrounded by a transport
boundary.  These islands are undesirable, since the goal of stirring
is to homogenize the entire domain.  Since the rod motion is periodic,
the islands can easily be identified using a Poincar\'e section
(stroboscopic map).  Our objective is to determine if the pair-loop
method can also find these islands, given relatively few trajectories.
We note that the pair-loop method would still apply if the system was
not time-periodic, but the Poincar\'e section would be
useless.

\begin{figure}
  \centering
  \includegraphics[width=0.5\figwidth]{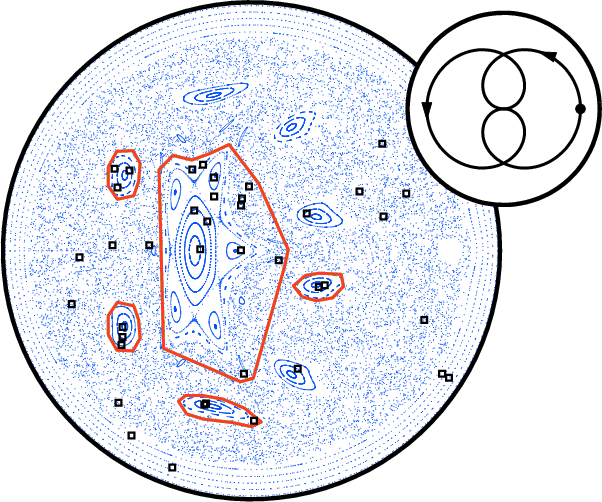}
  \caption{Poincar\'e section (stroboscopic map) for a vat of viscous
    fluid stirred by a rod (see inset for the rod's path).  A chaotic
    sea and several islands are visible, as well as regular orbits
    near the wall.  The squares indicate the initial position of
    the~$40$ trajectories, with non-growing loops detected by the
    pair-loop method drawn in.}
  \label{fig:trm}
\end{figure}

We specify an epitrochoidal trajectory of the rod as shown in the
inset to Fig.~\ref{fig:trm}. A set of~$40$ sample initial conditions
for the trajectories is also shown in the figure as small squares:
half of the initial conditions are distributed in seven islands, and
the other half throughout the well-mixed chaotic region.  The rod's
position is also taken as a trajectory, bringing the total to
forty-one strands in the braid.  The trajectories are tracked for five
periods of the rod's motion, and produce a sequence of seven thousand
generators.  The application of the pair-loop algorithm takes a few
minutes and produces loops roughly drawn in Fig.~\ref{fig:trm}.

The first observation is that not all of the islands have a
non-growing loop around them.  The reason for this is that there are
two islands which do not contain any trajectories, and two with
only one trajectory.  The algorithm relies on multiple trajectories
being inside a transport boundary in order to detect it.  If there are
no trajectories, then the method has no information about the region,
and a loop surrounding a single trajectory likewise provides no
information (and cannot even be encoded in Dynnikov coordinates).

The other observation is that some of the loops include nearby
trajectories in the chaotic sea.  The largest loop and the one at the
bottom of Fig.~\ref{fig:trm} are both examples of non-growing loops
that extend beyond the boundaries of the island.  The explanation for
this is that in the period of time analyzed none of the punctures in
the chaotic sea, the other islands, or the rod passes between the
island and its neighbors, which then end up being included in
the region.  If longer or more numerous trajectories were used, then
the probability that a puncture passes between the island and the
nearby puncture becomes higher, increasing the likelihood of detection
by the pair-loop method.

\section{Conclusions}
\label{sec:con}

In this paper we used tools from braid theory to find transport
boundaries using trajectory data in two-dimensional systems.  We
described how topological loops typically grow as a result of the
chaotic motion of the trajectories, and how this growth rate can be
related to a topological entropy.  The trajectory data is encoded as a
sequence of braid group generators, and the loops are represented by
Dynnikov coordinates.  We then described a method for detecting
non-growing loops, which enclose invariant regions.  In this
\emph{direct method} we simply try a very large number of loops and
look for the ones that grow slowly.

The direct method scales very poorly with the number of trajectories,
so we described a more complex approach, the \emph{pair-loop method},
where we examine a much smaller number of loops that connect pairs of
punctures.  Though algorithmically more involved, this method achieves
the same thing as the direct method, but at a much-reduced
computational cost.  The pair-loop method then allowed us to detect
regular islands in a simple rod-stirring device, using a set of~$40$
trajectories as well as the trajectory of the rod itself.

We discussed some of the limitations of the method in
Section~\ref{sec:rod}, but let us reiterate them here.  For the
detection method to be successful we need enough trajectories, but
also we need them to be long enough in time.  The number of
trajectories determines how well we can delimit invariant regions,
since we need at least two trajectories in a region to have a chance
at detecting it.  The length in time of the trajectories determines
whether loops have grown enough to be in their asymptotic growth
regime, so that we can distinguish them according to growth rate.

In practical applications, \LCSs\ often do not completely surround an
invariant region.  Thus, many regions tend to be quasi-invariant, but
for long times trajectories can escape and enter the region.  In our
method this would lead to an inexact determination of the structures,
or perhaps cause us to miss a quasi-invariant region altogether.  A
possible solution is to run trajectories for shorter times, which will
then make quasi-invariant regions show up as invariant regions.  For
example, in Fig.~\ref{fig:trm} some chaotic trajectories were ascribed
to the central region, even though they will eventually leak out.

Another potential issue is that the puncture (float) trajectories have
to be relatively accurate, especially if they happen to pass near each
other.  It is necessary to have the correct generator sequence in
order to find loops that do not grow.  If the position data does not
have a high enough resolution, it is possible that a crossing will be
missed, leading to inaccurate regions and false non-growing loops.
While it is straightforward to refine a trajectory from a velocity
field, this may not be an option for trajectory data taken from a
float in the ocean.

There are possible applications of this method beyond fluid dynamics.
Because the method depends only on trajectory information, the
trajectories need not be obtained from an incompressible fluid or even
a continuous medium.  For instance, one can search for clustering in
granular flow data, where the trajectories are now given by the motion
of individual grains, with no fluid between them.  In this case the
`invariant regions' correspond to clusters of granular particles that
evolve together.  This may allow a better understanding of transitions
in the state of a granular medium~\cite{Bonamy2002, Puckett2009,
  Lechenault2010}.  In future work we will apply the topological
detection method to this and other physical problems.

\section*{Acknowledgments}

The authors thank Matthew D.\ Finn for his help and insights, and for
the numerical simulation of the rod-stirring flow.  The authors also
thank George Haller and Tom Peacock for many helpful comments.  This
work originated at the Summer Program in Geophysical Fluid Dynamics at
the Woods Hole Oceanographic Institution, supported by NSF under grant
OCE-0824636.  MRA was supported by NSF under grant OCE-0645529, J-LT
under grant DMS-0806821.

\appendix

\section*{Appendix: Glossary of Terms Used}

\newcommand{\defemph}[1]{\textbf{#1}}

\begin{description}%\itemsep2pt

\item[braid, algebraic] an encoding of a physical braid in terms of
  braid group generators.  The generators are ordered temporally from
  left to right.

\item[braid, physical] a set of two-dimensional trajectories plotted
  in three dimensions, where the third dimension is time.

\item[braid group] the infinite group composed of all possible
  products of generators.

\item[braiding factor] sometimes used as a synonym for~$\Nint$ [see
  Eq.~\eqref{eq:int}], though originally it was defined in terms of
  the largest eigenvalue of the Burau
  representation~\cite{Thiffeault2005}, whose growth rate is only a
  lower bound for the growth of~$\Nint$~\cite{Fried1986, Kolev1989}.

\item[crossing number] the minimal number of times a topological loop
  intersects a given line.

\item[Dynnikov coordinates] a set of signed integers, obtained from
  crossing numbers, uniquely describing a topological loop.  For~$\nn$
  punctures, $2\nn-4$ integers are needed.

\item[entangled] a puncture is entangled by a given loop if the loop
  passes both above and below the puncture, with the canonical
  ordering of punctures from left to right.

\item[\EPS] the set of punctures that are each entangled by a given
  loop.  These are used in the pair-loop method to detect \IPSs.

\item[\ES] short for entangled puncture set.

\item[generators] after projecting a physical braid, each crossing is
  assigned a braid group generator, denoted~$\sigma_i$
  or~$\sigma_i^{-1}$, depending on whether the crossing is over-under
  or under-over.  For~$\nn$ punctures, $\nn-1$ generators are needed.

\item[invariant region] an ergodic component, in the language of
  dynamical systems.  Also called a dynamically-distinct region.
  Trajectories that initially start within such a region stay within
  the region, though the region itself may move and deform with time.
  An invariant region may be regular or chaotic.

\item[\LCS] a structure that can separate dynamically-distinct
  invariant regions.  (See \defemph{invariant region}.)

\item[link-loops] pair-loops connecting punctures within a given
  region.

\item[loop] a closed curve in the domain that encloses at least two
  punctures and does not intersect itself.  For our purposes, all
  loops are material loops, which means they move with the underlying
  flow.

\item[non-growing loop] a topological loop that surrounds an \IPS and
  has negligible growth when acted on by a sequence of generators.

\item[pair-loop] a loop or topological loop that encloses exactly two
  punctures.

\item[particle] an point in the domain, whose time-evolution gives a
  particle trajectory.  In our setting, they are the same as
  punctures.

\item[particle trajectory] see \defemph{trajectory}.

\item[puncture] a point in the domain that is a topological
  obstruction to loops.  See \defemph{trajectory}.

\item[region] see \defemph{invariant region}.

\item[slowly-growing loop] a topological loop that grows
  sub-exponentially when acted on by a sequence of generators.

\item[\IPS] the set of punctures that are enclosed
  by a non-growing loop.

\item[strand] a single trajectory in a physical braid.

\item[topological entropy (of a braid)] for a set of trajectories, the
  topological entropy can be thought of as the growth rate of a
  `rubber band' caught on the trajectories.  It is a measure of
  entanglement, as the length of the rubber band will grow faster if
  the trajectories frequently crisscross each other.  As the number of
  trajectories increases, the topological entropy of the braid
  converges to the \defemph{topological entropy of the flow}.  (See
  \defemph{braid, physical}.)

\item[topological entropy (of a flow)] this is the usual topological
  entropy from dynamical systems theory.  It measures the rate of
  loss of information about the precise identity of a trajectory.

\item[topological loop] same as a loop, but topological loops are
  equivalent if they can be deformed into each other continuously,
  without crossing the punctures.  We usually drop the word
  topological when the context makes it clear.  (In topology, these
  are known as equivalence classes of homotopic loops.)

\item[trajectory] a continuous path in the domain, parametrized by
  time.  At a given time, a trajectory is a puncture in the domain.
  (See \defemph{puncture}.)

\end{description}

%\bibliographystyle{siam}
%\bibliography{bib/journals_abbrev,bib/articles}


\end{document}